\documentclass[journal,comsoc]{IEEEtran}

\pdfoutput=1
\usepackage[T1]{fontenc}

\usepackage{cite}
\usepackage{amsmath,amssymb,amsfonts}
\usepackage{algorithmic}
\usepackage{graphicx}
\usepackage{subcaption}
\usepackage{mwe}
\usepackage{textcomp}
\usepackage{xcolor}
\usepackage{enumitem}
\usepackage[linesnumbered,ruled,vlined]{algorithm2e}
\usepackage{multicol}
\usepackage{multirow}
\usepackage{nomencl}
\makenomenclature
\usepackage{amsfonts}
\ifCLASSINFOpdf
\else
\fi
\usepackage{amsmath}
\usepackage[cmintegrals]{newtxmath}
\usepackage{float}

\begin{document}
%
\title{A Predictive Online Transient Stability Assessment with Hierarchical Generative Adversarial Networks
}

\author{Rui Ma,~\IEEEmembership{Student Member,~IEEE}, Sara Eftekharnejad,~\IEEEmembership{Senior Member,~IEEE}, Chen Zhong,~\IEEEmembership{Student Member,~IEEE}, Mustafa Cenk Gursoy,~\IEEEmembership{Senior Member,~IEEE}

}




\maketitle

\begin{abstract}
Online transient stability assessment (TSA) is essential for secure and stable power system operations. The growing number of Phasor Measurement Units (PMUs) brings about massive sources of data that can enhance online TSA. However, conventional data-driven methods require large amounts of transient data to correctly assess the transient stability state of a system. In this paper, a new data-driven TSA approach is developed for TSA with fewer data compared to the conventional methods. The data reduction is enabled by learning the dynamic behaviors of the historical transient data using generative and adversarial networks (GAN). This knowledge is used online to predict the voltage time series data after a transient event. A classifier embedded in the generative network deploys the predicted post-contingency data to determine the stability of the system following a fault. The developed GAN-based TSA approach preserves the spatial and temporal correlations that exist in multivariate PMU time series data. Hence, in comparison with the state-of-the-art TSA methods, it achieves a higher assessment accuracy using only one sample of the measured data and a shorter response time. Case studies conducted on the IEEE 118-bus system demonstrate the superior performance of the GAN-based method compared to the conventional data-driven techniques
\end{abstract}


\mbox{}
\nomlabelwidth=15mm 
\nomenclature{$D$,~$G$}{Discriminator and Generator of the GAN model}
\nomenclature{$r_{t}$}{The reset gate output of GRU}
\nomenclature{$z_{t}$}{The update gate output of GRU}
\nomenclature{$h_{t}^{'}$}{The current memory unit of GRU}
\nomenclature{$h_{t}$}{Final GRU output}
\nomenclature{$\sigma$}{Sigmoid activation function}
\nomenclature{$tanh$}{Hyperbolic tangent activation function}
\nomenclature{$x_{t}$}{Input data to GRU}
\nomenclature{$h_{t-1}$}{Output from the previous GRU}

\printnomenclature

\section{Introduction}

\IEEEPARstart{T}{ransient} 
stability is defined as the ability of all power grid synchronous machines to preserve synchronism after severe disturbances~\cite{alimi2020review}. 
The ever-increasing integration of inverter-based resources and the operation of power grids under more stressed conditions, has renewed the calls for fast online transient stability assessment (TSA) tools. An appropriate TSA will grant the system operators adequate time to take proper mitigative actions, such as intentional islanding or controlled load shedding.

The conventional TSA methodologies belong to one of three categories: 1) time-domain simulations, 2) direct methods, and 3) trajectory-based approaches. Time-domain methods solve a set of high-dimensional and nonlinear differential algebraic equations to assess the transient stability~\cite{fouad1991power,gurusinghe2015post}. These methods assume complete knowledge of the system model parameters and the operating conditions. However, their high computational complexity makes them impractical for near real-time applications. 
To enable faster transient stability assessment, direct methods, e.g., Lyapunov method~\cite{vu2015lyapunov,pai1989stability,farantatos2015predictive} and extended equal area criterion~\cite{xue1989extended,jahromi2016novel}, have been developed. The direct methods simplify TSA by energy functions that evaluate the dynamic performance of the system in a shorter time~\cite{pai2012energy,ju2017analytical}. Yet, because of the need to simplify the dynamic models, these methods do not scale well to large power systems.
The trajectory-based methods, such as Lyapunov exponents~\cite{yan2011pmu} and apparent impedance~\cite{li2014transient} methods, are relatively fast. However, at least a few cycles of post-fault transient data are needed for a precise stability assessment.

With the increased availability of high-resolution synchrophasor data, new TSA approaches have emerged where physical system constraints and system data inform each other for a more accurate TSA. Hence, without constructing a complete dynamic grid model, the dynamic stability status of power grids can be determined. 
Machine learning methods such as decision trees~\cite{diao2009design,amraee2013transient,li2014transient}, core vector machine~\cite{wang2016power}, support vector machine (SVM)~\cite{geeganage2014application}, and extreme learning machine~\cite{xu2012reliable,li2017application} have previously been deployed for TSA using system measurements, formulating TSA as a two-class (stable and unstable) classification problem. These data-driven methods require a conversion from raw Phasor Measurement Unit (PMU) data to selected features, which may result in the loss of critical dynamic information and adversely affect classification accuracy. 
Deep learning methods, such as convolutional neural network (CNN)~\cite{azman2020unified,zhu2019hierarchical}, long short-term memory (LSTM)~\cite{james2017intelligent,li2021fast}, gated recurrent unit (GRU)~\cite{pan2018stacked}, and the stacked denoising autoencoder (SDAE)~\cite{zhu2018deep} have been shown to address the problem of information loss by directly utilizing raw PMU data. 
However, to guarantee the assessment accuracy, longer durations of post-contingency time series data are needed, which will result in TSA delays. Longer delays are not desired for online TSA since the window of opportunity for taking an effective corrective action may be missed.

To address the aforementioned gaps in data-driven TSA methods, a novel data-driven method is developed in this paper to \textit{predict} the post-contingency PMU measurements that are used for classification. By predicting the data, shorter durations of post-contingency measurements are needed, while ensuring a high TSA accuracy.
The prediction capability is enabled by extending Generative Adversarial Networks (GAN), a model-free machine learning technique for generating datasets that closely mimic real datasets. Here, GAN enables learning the model behavior (system transient response in this case) to generate fictitious datasets. 

A GAN model consists of two neural networks: generative and adversarial networks. The generative network converts the input data drawn from a Gaussian distribution to synthetic data. The adversarial network classifies the generated synthetic data as either real or synthetic. The objective of the adversarial network is to maximize the difference between the real and synthetic data, while the generative model aims to minimize this difference~\cite{goodfellow2014generative}. GAN has been deployed in various domains for generating synthetic datasets including images~\cite{bao2017cvae} and music~\cite{mogren2016c}. In the context of power systems, GAN has been used to generate missing PMU data~\cite{ren2019fully} or address the lack of realistic PMU time series data~\cite{zheng2019synthetic,zheng2021generative,ruima_ias}.
Different from these applications that aimed at enhancing the historical data, in this paper, the GAN model is extended for \textit{predicting} the future PMU data. The prediction capability pursued by this work significantly differs from the previous applications. Hence, a different approach to adopting GAN is required to generate realistic future data and predict the system transient stability based on those data. This approach should: a) learn the spatial and temporal features of the transient data; b) predict the post-fault PMU data; and c) assess the transient stability of the system. The developed GAN-based TSA approach addresses these requirements for a fast and accurate stability assessment.

\begin{figure}
    \centering
    \includegraphics[width=\columnwidth]{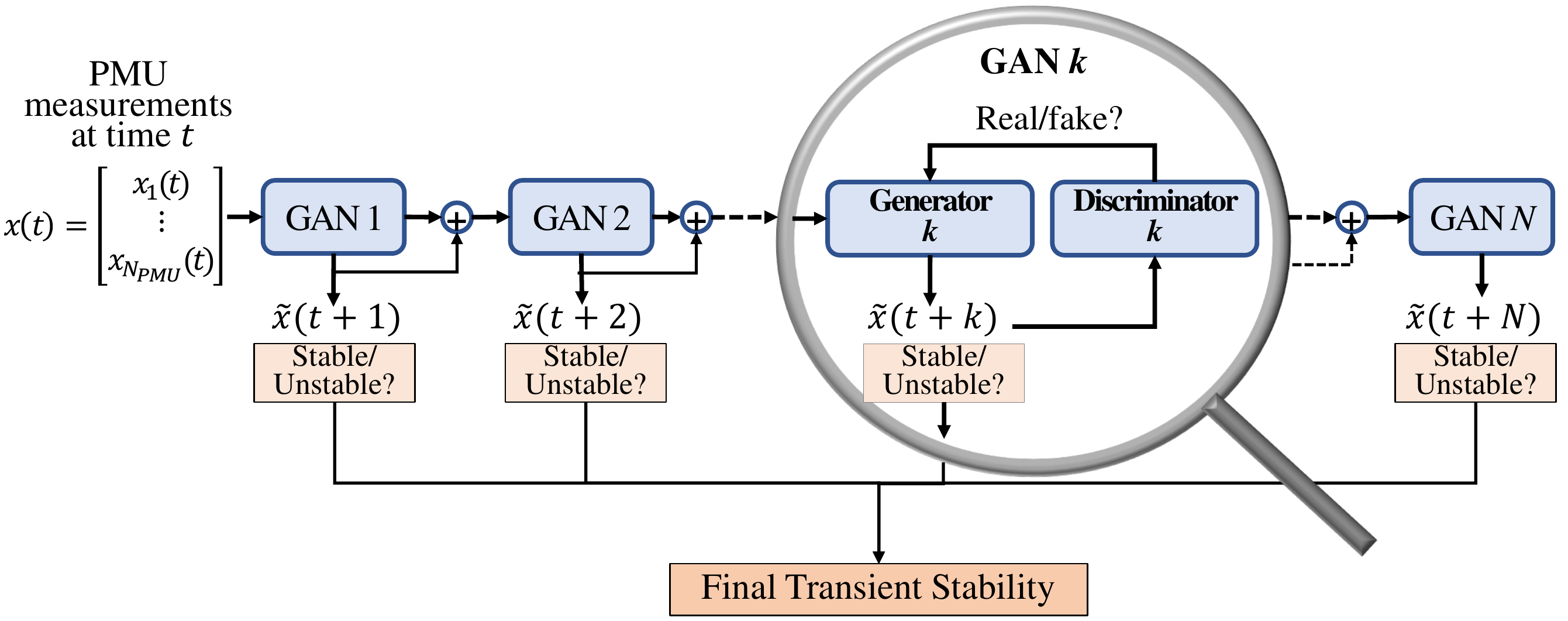}
    \caption{The structure of the developed HGAN-based TSA} \vspace{-3mm}
    \label{HGAN}
\end{figure}

The novelty of the developed TSA approach stems from the ability to rapidly and accurately determine transient stability, while only utilizing one sample of the measured post-contingency PMU data. This is achieved by developing a hierarchical structure, shown in Fig.~\ref{HGAN}, for the generative adversarial networks.
Several GAN models are stacked together to construct HGAN. The input to the hierarchical generative adversarial network (HGAN) is the measured post-contingency PMU data. The lowest level GAN utilizes the measured PMU data to predict the measurements for the next sampling time. The subsequent predictions are used by higher level GANs to further predict the PMU time series data. With the proposed structure, HGAN requires only a single sample of the measured data to predict the post-contingency time series transient data. 
In addition to the predicted sequence of the transient data, a binary classifier is embedded in the generative network of each sub-GAN. 
Each sub-GAN performs TSA individually, based on the measured one-sample PMU data as well as the predicted subsequent time series data from the lower-level sub-GANs.
The final TSA is achieved by combining the assessment results from all sub-GANs, with a goal to reduce the prediction error. 

The main contributions of this work are as follows:
\begin{enumerate}[leftmargin=*]
    \item A hierarchical GAN-based TSA is developed to accurately predict the transient status after a disturbance. Only a single sample of post-contingency PMU voltage data is needed. Hence, TSA can be performed in a relatively short period of time;
    \item The generative and adversarial networks are designed in a manner to retain the spatial and temporal features of the multivariate PMU time series data and improve TSA accuracy;
    \item Case studies are conducted to demonstrate the ability of the developed model in predicting a sequence of the transient data. In comparison to the conventional machine learning-based TSA methods that deploy decision trees, SVM, LSTM, or GRU, the developed model achieves a higher TSA accuracy with only one sample of PMU measurements.
\end{enumerate}

The rest of this paper is organized as follows: Section II briefly describes the concept of transient stability and formulates the TSA problem. Section III introduces the conventional GAN model and illustrates each element of the developed HGAN model. The developed HGAN model is tested on the IEEE 118-bus system in Section IV. The conclusions are provided in Section V.

\section{Transient Stability Assessment}
During a power system disturbance, such as a transmission line fault, large swings appear in generator rotor angles, bus voltages, or line currents~\cite{zhu2018deep}. To ensure a stable grid, all generators should maintain synchronism after the clearance of the faults. If the synchronism is lost, prompt mitigations such as controlled islanding need to be taken to prevent further failures. This remarks the need for an online TSA approach that can promptly and precisely identify the transient status at the earliest stages of a fault clearance.

\begin{figure}
    \centering
    \includegraphics[width=\columnwidth]{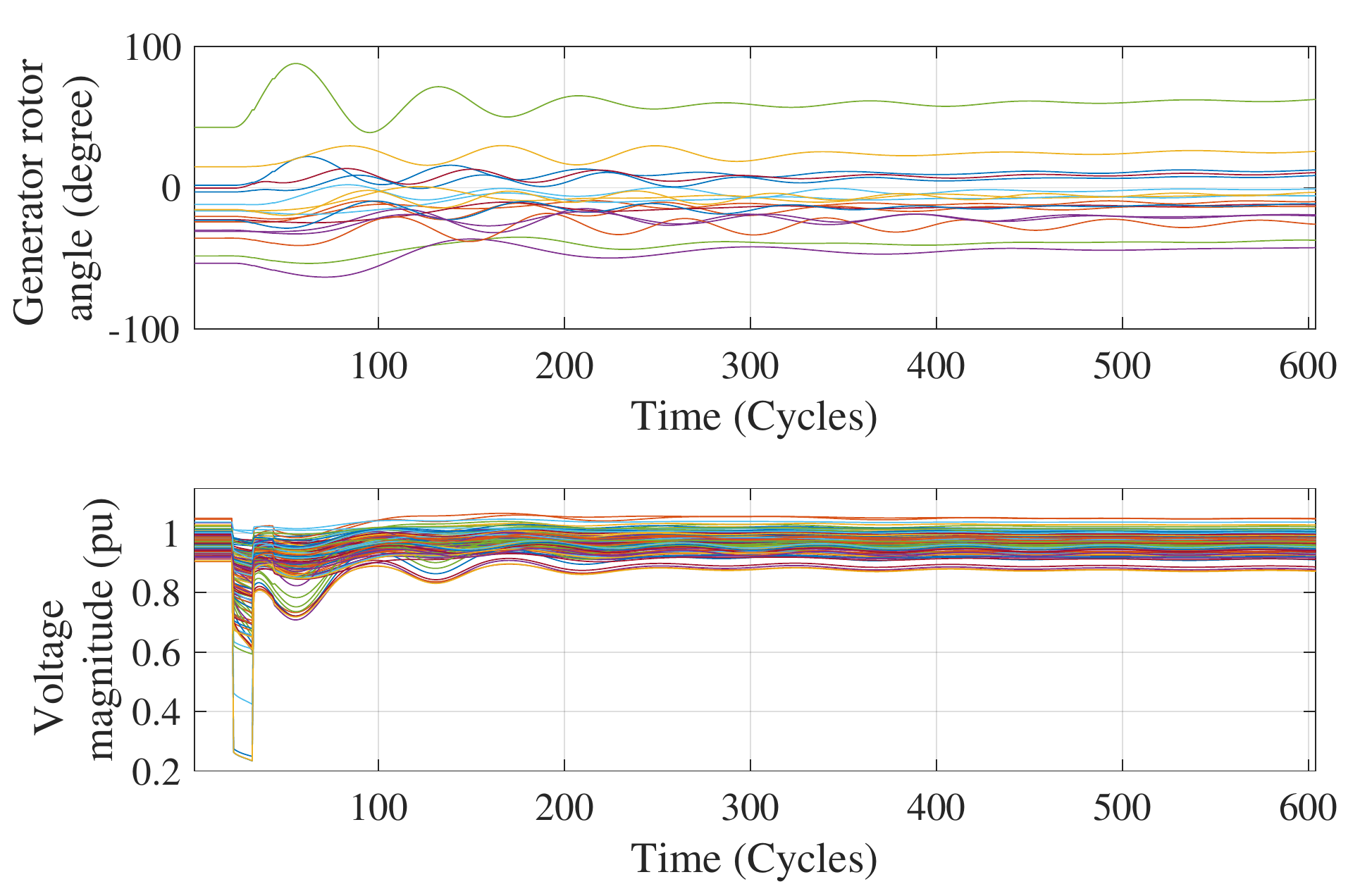}
    \caption{The trajectories of the generator rotor angles and voltage magnitudes in the IEEE 118-bus system. A three-phase short circuit is simulated on line 102 at 0.167s and is cleared after 5 cycles (0.083 s).  } \vspace{-3mm}
    \label{transient}
\end{figure}

\begin{figure}[]
        \centering
        \begin{subfigure}[b]{0.485\columnwidth}
            \centering
            \includegraphics[width=0.5\columnwidth]{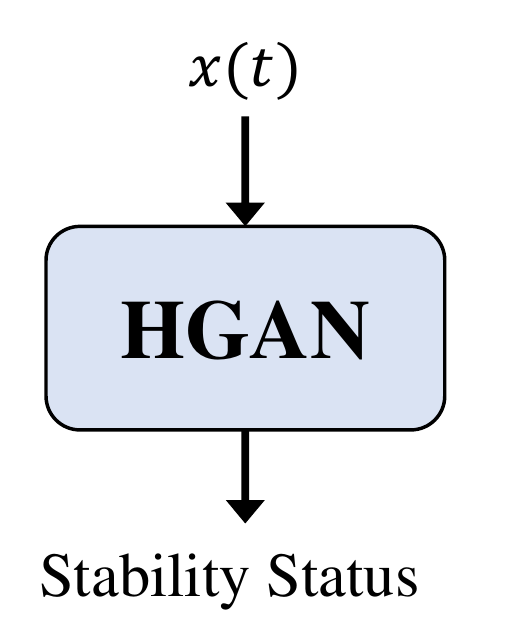}
            \caption[]%
            {{\small The developed HGAN-based TSA method}}    
            \label{HGAN OBJ}
        \end{subfigure}
        \hfill
        \begin{subfigure}[b]{0.485\columnwidth}  
            \centering 
            \includegraphics[width=0.5\columnwidth]{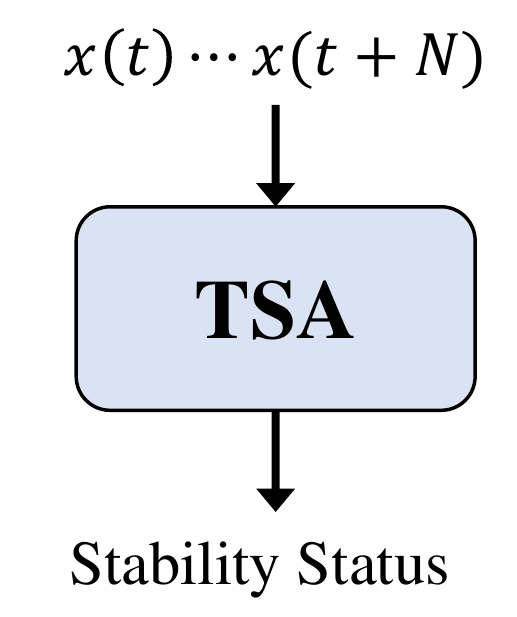}
            \caption[]%
            {{\small The conventional machine learning-based TSA methods}}  
            \label{TSA OBJ}
        \end{subfigure}
        \caption[ ]
        {\small The input and output of the developed HGAN model and the conventional machine learning-based TSA methods. $x(t)$ is the first sampled post-contingency PMU data. The number of PMU samples is $N$.} \vspace{-3mm}
        \label{objective}
\end{figure}

Fig.~\ref{transient} depicts the post-fault dynamics of the generator rotor angles and bus voltages in the IEEE 118-bus system. A three-phase short circuit is simulated on line 102 at time instant 0.167s, which is cleared after 5 cycles, and Generator 1 is the reference machine. As seen, large excursions occur in the trajectories of the generator rotor angles and bus voltage magnitudes when the described fault occurs. 
To classify the system transient stability status, a commonly used stability index based on rotor angles is used~\cite{pavella2012transient}:
\begin{equation}\label{transient metric}
    \eta = \frac{360^{\circ}-\Delta \sigma_{max}}{360^{\circ}+\Delta\sigma_{max}}
\end{equation}
where $\Delta \sigma_{max}$ is the absolute value of the maximum post-fault angle difference between any two generators. The stability is defined as,
\begin{equation}
    \delta = \left\{\begin{matrix}
\text{stable;} & \eta > 0\\ 
\text{unstable;} & \eta \leq 0.
\end{matrix}\right.
\end{equation}
In the transient event depicted in Fig.~\ref{transient}, $\eta=41.3$. Therefore, the IEEE 118-bus system is considered stable after clearing the three-phase fault on line 102. 
However, using~(\ref{transient metric}) for determining the transient status requires the estimation of $\Delta \sigma_{max}$ over a relatively long time (a few seconds), which is not desired for online TSA.
Hence, the objective of the developed TSA approach, illustrated in Fig.~\ref{HGAN OBJ}, is to acquire the transient status at the earliest stages of a contingency.  
As illustrated in Fig.~\ref{TSA OBJ}, unlike the other machine learning-based TSA methodologies that require a large volume of post-contingency PMU data, the developed method only requires the first sample of the PMU data after the clearance of the faults. Thus, more time can be granted for finding the optimal corrective actions. 
In this work, the PMU voltage data are selected for TSA for two main reasons: (a) the voltage phasors are readily available from the PMU measurements; (b) previous TSA studies have shown that bus voltage data can be utilized to accurately predict the transient stability~\cite{james2017intelligent,zhu2018deep,del2007estimation}.
It is also shown in Section~\ref{analysis_results} that only utilizing PMU voltage magnitude data is sufficient to achieve a high assessment accuracy. However, other PMU measurements that contain the system response, such as bus voltage angles and line current phasors, can also be deployed by the developed HGAN-based TSA method. Additionally, to mimic real power grids, only some buses are equipped with PMUs (e.g., 20 buses in the IEEE 118-bus system). The PMUs are added in a manner to ensure full system observability.


\section{Predictive Online Transient Stability Assessment}
A new approach is developed for near real-time transient system assessment, which is fast, accurate, and requires significantly less PMU data than the conventional techniques. The developed approach is twofold: 1) predicting the post-contingency PMU voltage time series data, and 2) assessing the transient stability using the predicted data. In this section, the fundamentals of the GAN are introduced first, followed by a detailed description of each element of the HGAN-based TSA method.

\subsection{The Generative Adversarial Network}
GAN is an unsupervised learning approach that has been used in many applications for generating synthetic images, music, time series data, among others~\cite{yoon2019time,radford2015unsupervised,bao2017cvae,mogren2016c,frid2018gan}. GAN learns the distribution of a real dataset and generates synthetic data that are close to the real data~\cite{goodfellow2014generative}. In the context of power systems, GAN-based methods have been developed to recover missing PMU data~\cite{ren2019fully} and generate synthetic PMU datasets~\cite{zheng2019synthetic,zheng2021generative,ruima_ias}.

GAN consists of two deep neural networks: generator $G$ and discriminator $D$. The generator generates synthetic data while retaining the statistical features of the real data. The discriminator differentiates the real data from the fictitious data generated by the generator. The two networks $G$ and $D$ are trained iteratively until a Nash equilibrium is achieved, which indicates that the generated synthetic data and real data cannot be distinguished~\cite{ren2019fully}.

For a random Gaussian noise space $\mathbf{Z}$, the generator maps the input noise $z$ drawn from $\mathbf{Z}$ to a synthetic dataset $\hat{x}$ as,
\begin{equation}\label{G}
    \mathbf{G}(z | \theta_{g}): z \rightarrow \hat{x}
\end{equation}
where $\theta_{g}$ is the neural network parameters of $G$. The objective of the generator is to map a random noise to synthetic data, such that the distribution of the synthetic data $P(\hat{x})$ is close to the real data $P(x)$.
The discriminator $D$ estimates the probability, $D(\cdot|\theta_{d})$, that the input data is real rather than synthetic, where $\theta_{d}$ the is the neural network parameters of $D$. 
The discriminator aims to maximize the difference between the real and synthetic data so that they could be distinguished,
\begin{equation}\label{D max}
    \begin{split}
        \underset{D}{\text{Max}} ~V(G,D)=& \mathbb{E}_{x\sim P(x)}[\log D(x)] \\&+\mathbb{E}_{z\sim P(z)}[\log (1-D(\hat{x})]
    \end{split}
\end{equation}
where $V(\cdot)$ is the loss function. 
The first term $\mathbb{E}_{x\sim P(x)}[\log D(x)]$ is the probability that the discriminator classifies the real data as real. The second term is the probability that the synthetic data $\hat{x}$ generated from $G$ is classified as fake by $D$.
The objective of $G$ is to minimize the difference between the real and synthetic data,
\begin{equation}
    \underset{G}{\text{Min}} ~V(G,D) = \mathbb{E}_{z\sim P(z)}[\log (1-D(\hat{x}))].
\end{equation}
As the generator and discriminator are trained together, the objective of GAN is to address a minimax problem:
\begin{equation}\label{GAN minimax}
    \begin{split}
        \underset{G}{\text{Min}}\underset{D}{\text{Max}}~V(G,D)=&\mathbb{E}_{x\sim P(x)}[\log D(x)]
        \\ &+ \mathbb{E}_{z\sim P(z)}[\log (1-D(\hat{x}))].
    \end{split}
\end{equation}
The GAN training algorithm is described in Algorithm~\ref{GAN training algorithm}. In each training episode, by leveraging the stochastic gradient descent algorithm, the generator and discriminator are updated, such that the synthetic data are closer to the real data. 

\begin{algorithm}[t]
\SetAlgoLined
 Initialize the neural network parameters of the generator and discriminator, i.e., $\theta_{g}$ and $\theta_{d}$ respectively\;
 \For{$ep=1,\cdots,ep_{max}$}{
  Obtain a minibatch of the random noise $z$ from $P(z)$, and another minibatch of the real data $x$ from $P(x)$\;
  Generate the synthetic data $\hat{x}=G(z|\theta_{g})$\;
  Feed the generated synthetic data $\hat{x}$ and the real dataset $x$ into the discriminator $D$, and obtain the estimated probability $D(x|\theta_{d})$ and $D(\hat{x}|\theta_{d})$\;
  Update $G$ and $D$ by descending their gradients:
  \begin{equation*}
  \begin{split}
      &\triangledown _{\theta_{g}}(\mathbb{E}_{z\sim P(z)}[log(1-D(\hat{x}))])
      \\&\triangledown _{\theta_{d}}(\mathbb{E}_{x\sim P(x)}[logD(x)]+\mathbb{E}_{z\sim P(z)}[log(1-D(\hat{x})])
  \end{split}
  \end{equation*}\vspace{-5mm}
 }
\caption{The GAN training steps}\label{GAN training algorithm}
\end{algorithm}


\subsection{Gated Recurrent Units for Feature Extraction from PMU Time Series Data}
When predicting the future dynamic response in a power system based on data, it is critical to preserve the spatial and temporal features of the time series data to ensure accuracy. Gated Recurrent Units (GRUs), a variant of RNN, enable learning the spatial and temporal features of time series data~\cite{zaremba2014recurrent}. Thus, the developed TSA approach leverages GRU to construct the generator and the discriminator. 

\begin{figure}
    \centering
    \includegraphics[width=0.6\columnwidth]{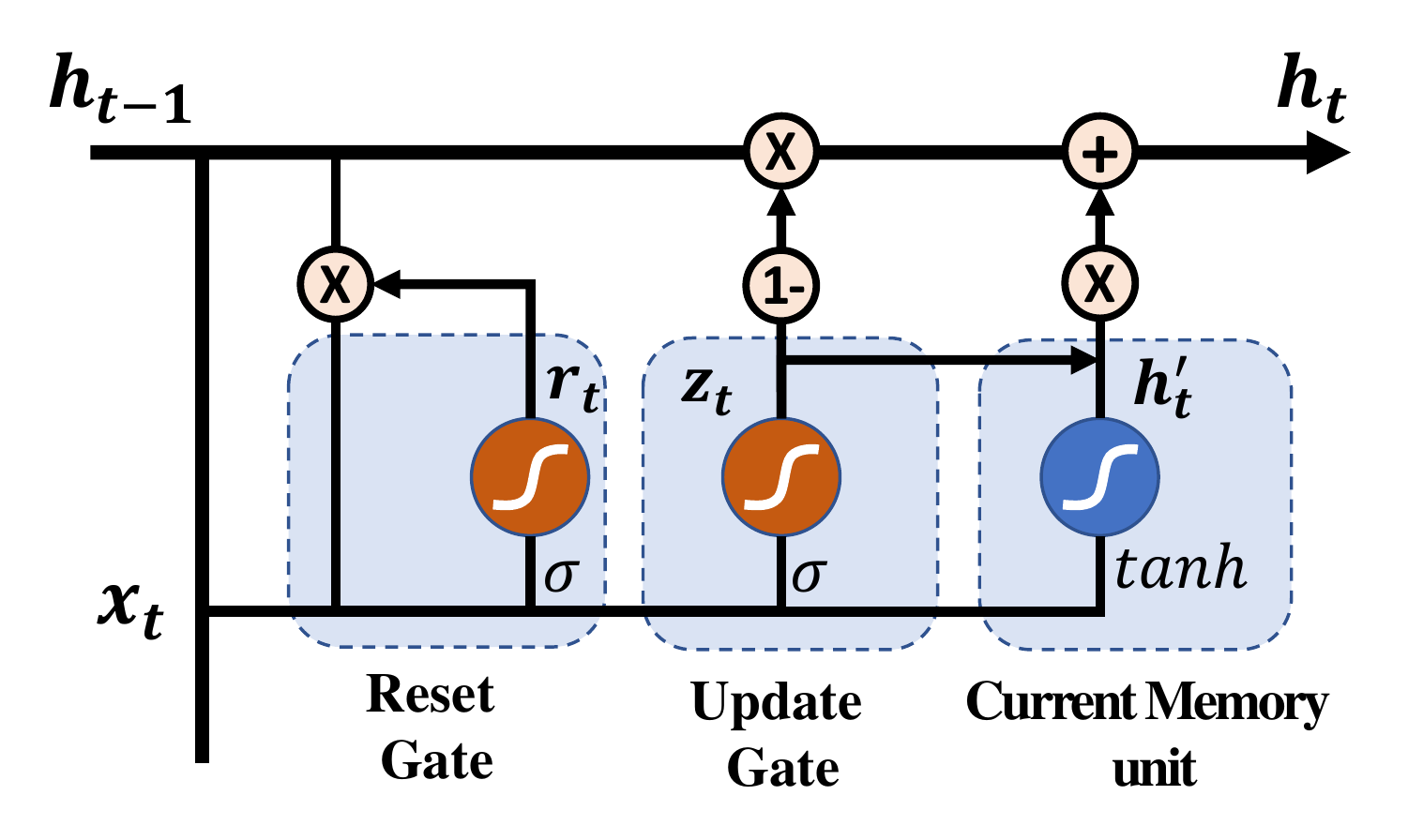}
    \caption{The structure of GRU} \vspace{-5mm}
    \label{GRU}
\end{figure}

Equipped with directed cycles in network connections, RNN retains information from past data, so that it can be combined with the present data to more accurately predict the future sequence of data.
However, RNN suffers from the vanishing or exploding gradient problem during training~\cite{pascanu2013difficulty}. If the RNN is not designed properly, the gradient may be exponentially updated towards zero, referred to as a vanishing gradient, or exponentially diverge to infinity, referred to as an exploding gradient. Either of these problems challenges the proper RNN training.
To resolve this problem, GRU is equipped with a specific mechanism to determine the volume of past data, to be added to the output. By doing so, the problem of vanishing or exploding gradient is eliminated. As illustrated in Fig.~\ref{GRU}, GRU consists of three elements: the reset gate, the update gate, and the current memory unit~\cite{zaremba2014recurrent}. The update gate determines the volume of past data that are used for predicting the future data sequence, while the reset gate determines the volume of the stored past information to be forgotten. The current memory unit uses the output from the reset gate to store the relevant information from the past data. The output of GRU is the sum of the output from the three gates.
The three gates and the final output can be described as, 
\begin{subequations}\label{GRU function}
\renewcommand{\theequation}{\theparentequation.\arabic{equation}}
\begin{gather}   
r_{t} = \sigma ( W^{r}x_{t} + U^{t}h_{t-1} )
\\
z_{t} = \sigma (  W^{z}x_{t} + U^{z}h_{t-1} )
\\
h_{t}^{'} = \tanh (Wx_{t} + r_{t}\odot Uh_{t-1})
\\
h_{t} = z_{t}\odot h_{t-1} + (1-z_{t})\odot  h_{t}^{'} .
\end{gather}
\end{subequations}
In~(\ref{GRU function}), $W$ and $U$ are the weights of the GRU network, and $\odot$ is the Hadamard (element-wise) product. 


\subsection{The Developed HGAN-based TSA Method}\label{ensemble section}
The developed HGAN-based TSA aims to predict the post-fault transients and assess the system's transient stability. 
The developed method also improves the GAN structure by learning the spatial and temporal features of the multivariate PMU time series data. As deploying one type of PMU data is sufficient to assess the transient stability, the PMU voltage magnitude data are utilized for TSA in this paper. However, other PMU measurements may also be used by the HGAN-based TSA model. The measured post-fault PMU voltage magnitudes are the inputs to the GAN model, i.e., $x(t)$. The data for the next time instant, i.e., $\hat{x}(t+1)$ are predicted by the aforementioned GAN structure. However, the measured data $x(t)$ and the predicted data $\hat{x}(t+1)$ may not have sufficient information to determine the transient status. To address this deficiency, it is proposed in this paper to stack multiple GANs to predict a longer sequence of future data to better reveal the transient status of the system. As a result, a higher transient assessment accuracy can be achieved. 

The structure of the developed HGAN-based TSA is shown in Fig.~\ref{HGAN}. Assuming that HGAN has $N$ stacked GANs, the measured PMU post-fault voltage data is fed into the lowest level GAN, i.e., \textit{GAN 1} to predict the data for the next time instant $\hat{x}(t+1)$. The predicted data is combined with the measured data $x(t)$ and fed into \textit{GAN 2} for subsequent predictions. Therefore, the HGAN-based TSA with $N$ GANs can generate $N+1$ voltage time series data, $\{x(t), \hat{x}(t+1),\cdots,\hat{x}(t+N)\}$. 

A binary classifier is embedded in each generator of the stacked GANs for TSA classification. In each \textit{GAN $k$}, the embedded classifier utilizes the input sequence data $\{x(t),\hat{x}(t+1),\cdots,\hat{x}(t+k-1)\}$ to determine the TSA classification, i.e., stable or unstable. The classification results from all sub-GANs are then combined by using an average-based ensemble strategy to determine the final transient stability assessment. The generator has three objectives: 1) learning the distribution of the real measurements; 2) predicting the next sequence of the time series data; 3) assessing the transient stability status. Based on these objectives, the generator loss function in \textit{GAN $k$} is formulated as,
\begin{equation}\label{g loss in HGAN}
    \begin{split}
        \mathbf{L}_{G_{k}} = &\log(1-D(\hat{x}(t+k))) + ( \hat{x}(t+k) - x(t+k) )^{2} 
        \\&+ y\log(p(y)) + (1-y)\log(1-p(y))
    \end{split}
\end{equation}
where $y$ is the binary label of the transient data. Here, a zero label is assigned to unstable transients, and one to stable transients.
The probability that the data is labeled as stable is defined by $p(y)$. 
In formulating the loss function in~(\ref{g loss in HGAN}), the square of the error, $( \hat{x}(t+k) - x(t+k) )^{2}$, is used to minimize the error of the predicted data $\hat{x}(t+k)$. The cross entropy loss $y\log(p(y)) + (1-y)\log(1-p(y))$ is included so that the generator correctly classifies the system stability status. 
In addition, the GRU structure is used to construct generative and adversarial neural networks to better learn the spatial and temporal features of the time series data.

\begin{figure}
    \centering
    \includegraphics[width=0.9\columnwidth]{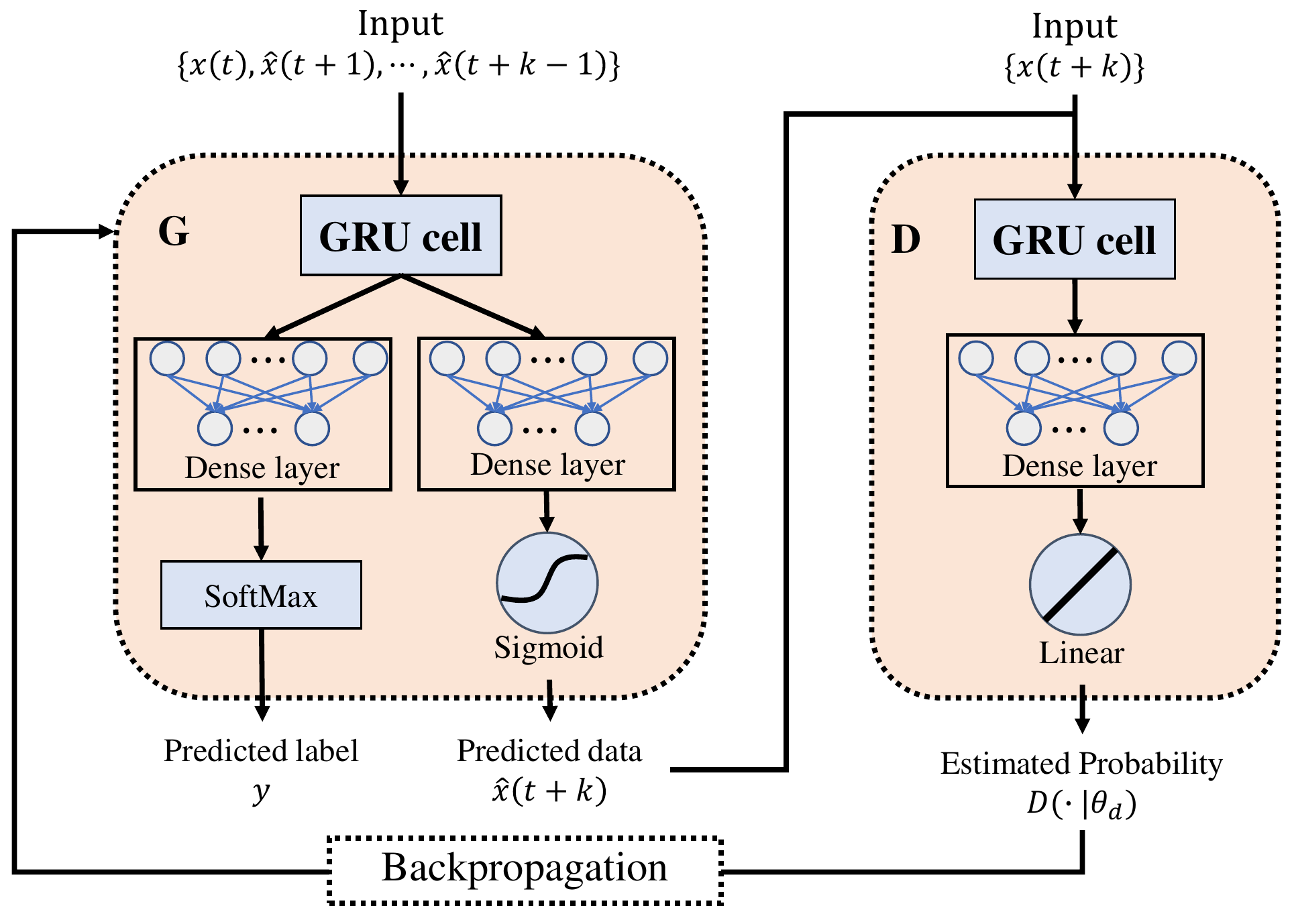}
    \caption{The structure of  GAN $k$ in the HGAN-based TSA}\vspace{-3mm}
    \label{sub-GAN detail}
\end{figure}

The detailed structure of \textit{GAN $k~(k\in\{0,1,\cdots,N\})$} is illustrated in Fig.~\ref{sub-GAN detail}. 
The input to \textit{GAN $k$} is the measured and predicted PMU voltage data $X = \{x(t),\hat{x}(t+1),\cdots,\hat{x}(t+k-1)\},~X\in\mathbf{R}^{k\times N_{PMU}}$, where $N_{PMU}$ is the total number of PMUs deployed in the system. 
In the generator $G$, the input $X$ is first fed into the GRU cell to learn the spatial and temporal features of the time series data. To improve the learning efficiency of the generator, the fully connected layer is added after the GRU cell. As the generator has two different outputs, i.e., the predicted label and subsequent data, two parallel fully connected layers are used to perform these two tasks separately. Furthermore, to generate the predicted label $y$ and the predicted data $\hat{x}(t+k)$, the \textit{softMax} and \textit{Sigmoid} activation functions are deployed to interpret the fully connected layer output.
The discriminator $D$ also uses GRU to construct the neural networks. The output of the GRU in the discriminator is sent to the fully connected layer, followed by a linear activation function that converts the output to the estimated probability of the input data $D(\cdot |\theta_{d})$. The estimated probability is then backpropagated to the generator to update the parameters of the generative network.

The prediction accuracy of the upper level GAN depends largely on the accuracy of the lower GANs. A small prediction error from a lower level GAN will propagate to the higher levels and eventually lead to a large prediction error. Hence, in this study, the GANs are trained sequentially. For instance, when training \textit{GAN $k$}, the network parameters of the well-trained lower GANs $i,~(i\in\{0,\cdots,k-1\})$, are kept fixed. 
Otherwise, repeatedly updating the well-trained networks of the lower GANs when training the upper GANs will cause an overfitting problem, and thus reduce the prediction accuracy.
The training process of the HGAN-based TSA is given in Algorithm~\ref{HGAN training algorithm}. For each GAN model, deploying the stochastic gradient descent algorithm, the generator and the discriminator are updated until the convergence of the cross-entropy loss, or until the training episode reaches its maximal $ep_{max}$, whichever occurs sooner.

\begin{algorithm}[t]
\SetAlgoLined
    Normalize the data such that the maximum normalized voltage is one\;
  \For{$i=1,\cdots,N$}{
    Initialize the neural network parameters of the generator and discriminator, i.e., $\theta_{g}$ and $\theta_{d}$ in GAN $i$, respectively \;
     \For{$ep=1,\cdots,ep_{max}$}{
      Obtain a minibatch of the measured post-fault PMU voltage data $x(t)$ and $x(t+i)$\;
      Using the well-trained lower GANs $k,~k\in\{1,\cdots,i-1\}$ to obtain the sequence data $\{x(t),\hat{x}(t+1),\cdots,\hat{x}(t+i-1)\}$\;
      Feed the obtained sequence data to the generator $G$, generate the data label $y$, and predict data $\hat{x}(t+i)$ for the next sampling time\;  
      Input the predicted data $\hat{x}(t+i)$ and real data $x(t+i)$ to the discriminator $D$, and obtain the estimated probability $D(\hat{x}(t+i)|\theta_{d})$ and $D(x(t+i)|\theta_{d})$\;
      Update the generator and discriminator with gradient descent algorithm;
 }
 Save the network parameters $\theta_{g}$ and $\theta_{d}$ for GAN $i$\;
  }
\caption{The training algorithm for the developed HGAN-based TSA}\label{HGAN training algorithm}
\end{algorithm}
The trained HGAN-based TSA can be applied online for stability assessment. Once the first post-fault PMU voltage data, i.e., $v(0)$, is measured, the HGAN-based model with $N$ sub-GANs generates the subsequent voltage data $\{v(0),\hat{v}(1),\cdots,\hat{v}(N)\}$. The TSA results from these $N$ sub-GANs will be combined by using an average-based ensemble technique for the final TSA classification. Here, it is assumed that the final TSA result is stable if more than half of sub-GANs vote for stable; otherwise, the system is unstable. The use of the ensemble method reduces the variance of the prediction errors from all sub-GANs
and results in a more accurate and reliable TSA~\cite{hansen1990neural}.

\section{Case Studies}
To evaluate the performance of the HGAN-based TSA under various system operating conditions, the IEEE 118-bus system with 118 buses, 170 lines, and 9 transformers is studied. To demonstrate the effectiveness of the HGAN-based TSA method, four different baseline methods, namely, decision trees, SVM, LSTM, and GRU are used for comparison purposes. Furthermore, as the TSA accuracy of the HGAN-based method relies on PMU measurements, sensitivity studies are performed to investigate the impact of noise,  number, and location of PMUs on TSA accuracy. The simulations are conducted on an i7 computer with a 3.2GHz CPU and 64GB RAM.

\subsection{PMU Data Generation}
The post-fault PMU voltage time series for testing and training are generated with simulations.
Various operating conditions (90\%, 100\%, and 110\% of the nominal loading condition) are simulated under different system typologies, i.e., normal and one line out of service (due to maintenance). The contingencies considered in this study are three-phase faults on buses and transmission lines (located at 30\%, 50\%, and 80\% of the lines). The faults are cleared after 5 cycles, and the system topology is kept unchanged after clearing the fault. To minimize the adverse impacts of an imbalanced dataset on the accuracy of the HGAN-based TSA, 5,000 stable and 5,000 unstable transients are generated. Furthermore, 80\% of the data are used for training, and the remaining 20\% are utilized for testing. For creating stability labels, the transient stability of each event is determined using the index outlined in~(\ref{transient metric}).
To generate the PMU time series data, the sampling frequency of the simulation is set to 120 frames per second. Additionally, to mimic real power grids where not every bus is equipped with PMUs, 20 PMUs are deployed in the IEEE 118-bus system to ensure full observability~\cite{ma2020pmu}. The locations of the PMUs are given in Table~\ref{PMUplacement}. The training and testing data can be found at the IEEE DataPort (https://dx.doi.org/10.21227/6f5v-q924).
\begin{table}[]
\centering
\caption{PMU placement in the IEEE 118-bus system} 
\label{PMUplacement}
\resizebox{0.9\columnwidth}{!}{%
\begin{tabular}{c|c}
\hline \hline
\textbf{Test System} & \textbf{PMU placement} \\ \hline
\textbf{IEEE 118-bus system} & \begin{tabular}[c]{@{}c@{}}5, 12, 15, 17, 32, 37, 49, 56,\\ 59, 67, 69, 70, 71, 77, 80, 85,\\ 92, 96, 100, 105\end{tabular} \\ \hline
\end{tabular}}\vspace{-1mm}
\end{table}


\subsection{Performance Analysis of the HGAN-based TSA}\label{analysis_results}

\noindent \textbf{---Performance Analysis}:
Upon obtaining the training and testing datasets that contain transient events, each level GAN of the developed TSA model is trained, following the training steps outlined in Algorithm~\ref{HGAN training algorithm}. The training parameters are provided in Table~\ref{HGAN parameters}. It is found that the HGAN-based model with three GANs is sufficient for an accurate stability assessment. Although more GANs generate longer duration of voltage time series data, and thus better represent a transient pattern, a small error in lower level GANs could propagate through the stacked GANs, and eventually lead to a large cumulative error. The learning rates of the generator and discriminator are set to 0.001 and 0.0001, respectively. A larger learning rate can increase the learning speed, while the results may converge to a suboptimal solution. Additionally, to address the overfitting problem during the training of a GAN, the neural network parameters of the lower level GANs are not updated.
\begin{table}[]
\centering
\caption{Parameters of the HGAN-based TSA model}
\label{HGAN parameters}
\resizebox{0.9\columnwidth}{!}{%
\begin{tabular}{c|c}
\hline \hline
\textbf{Parameters} & \textbf{Value} \\ \hline
Number of GANs & 3 \\ \hline
Number of episode & 20000 \\ \hline
Learning rate of generator / discriminator & 1e-3 / 1e-4 \\ \hline
Number of GRU layers /  hidden layers & 2 / 30 \\ \hline
Batch size & 128 \\ \hline
\end{tabular}%
} 
\end{table}

\begin{figure*}[]
\begin{multicols}{3}
    \includegraphics[width=\linewidth]{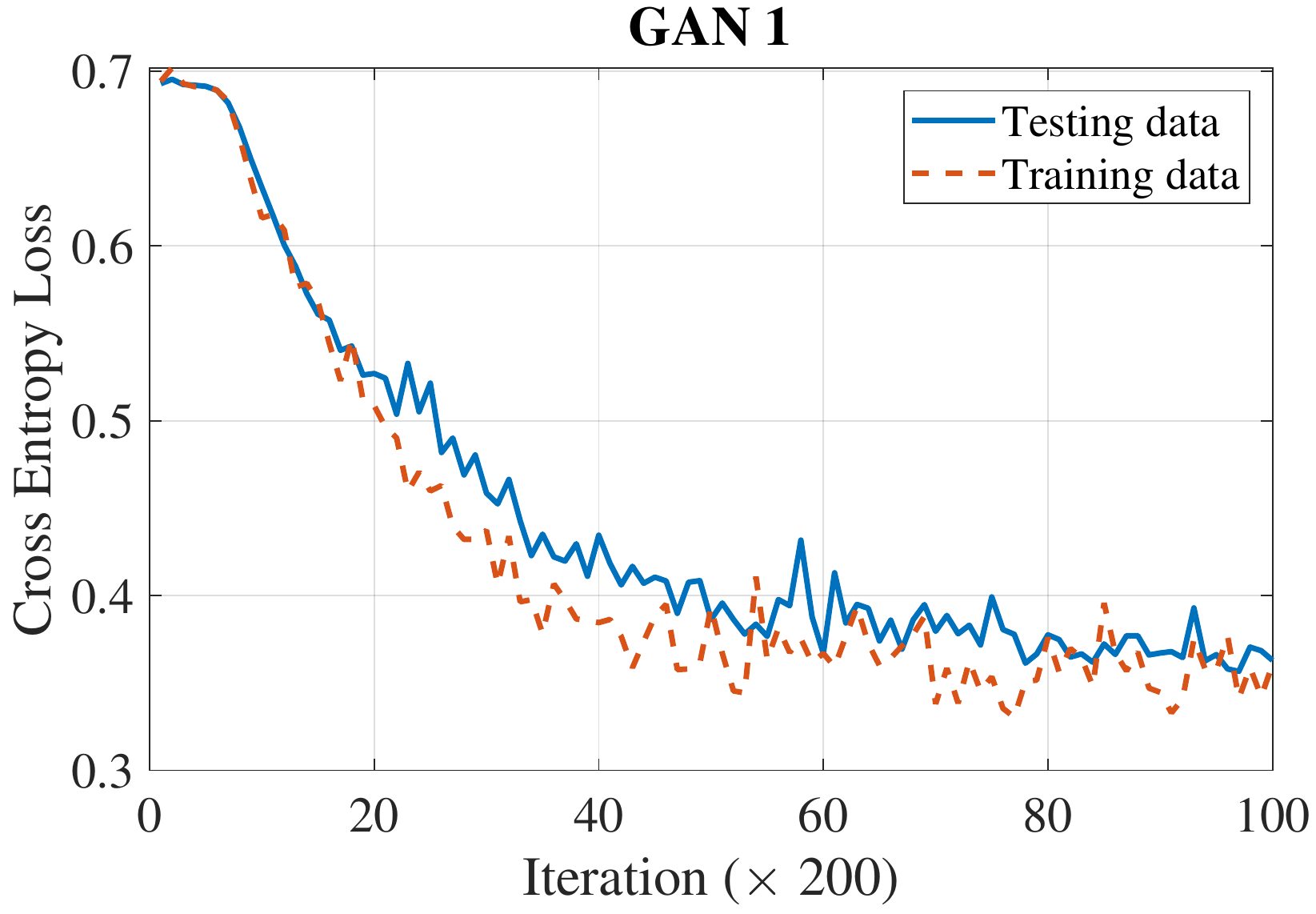}\par 
    \includegraphics[width=\linewidth]{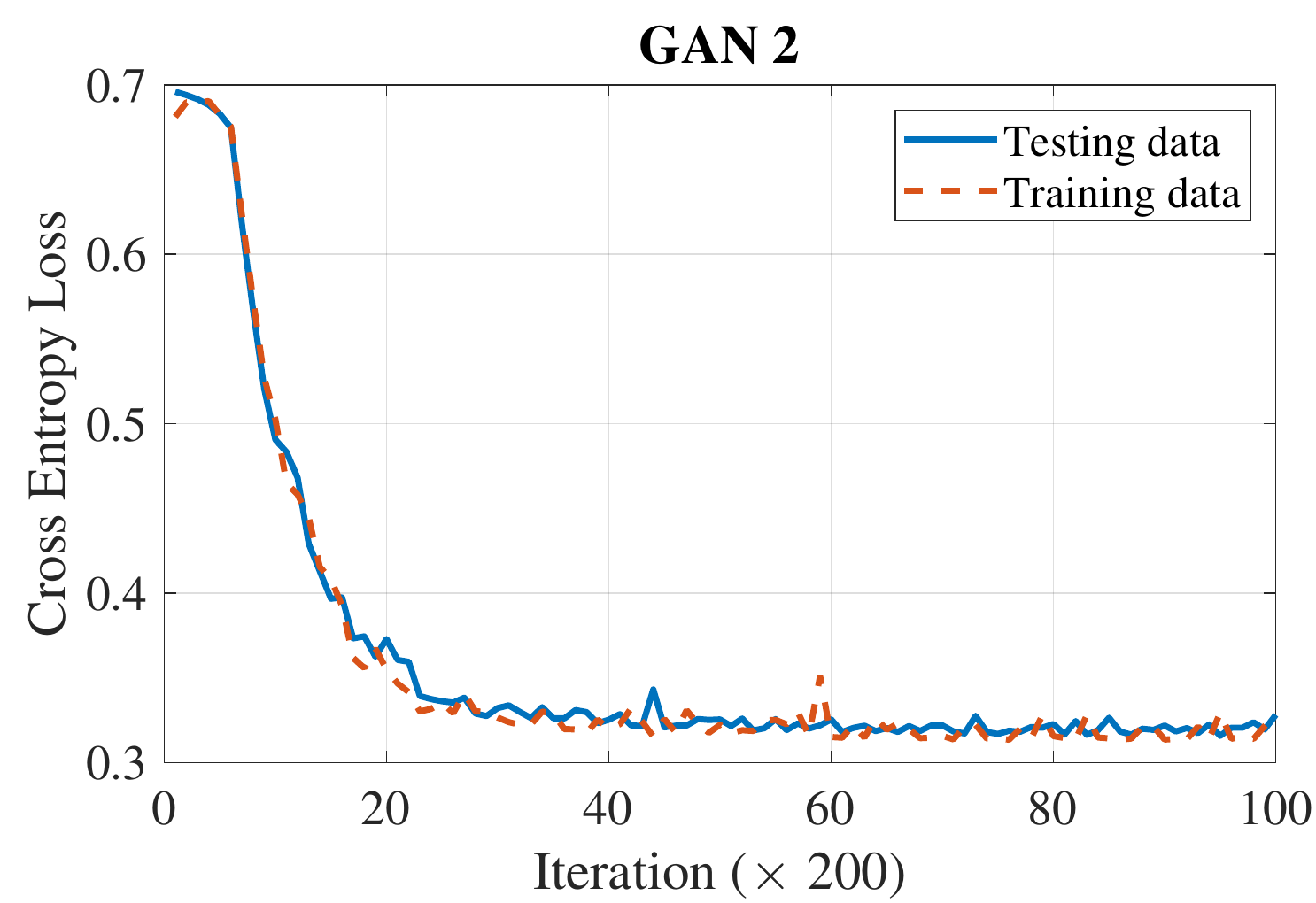}\par 
    \includegraphics[width=\linewidth]{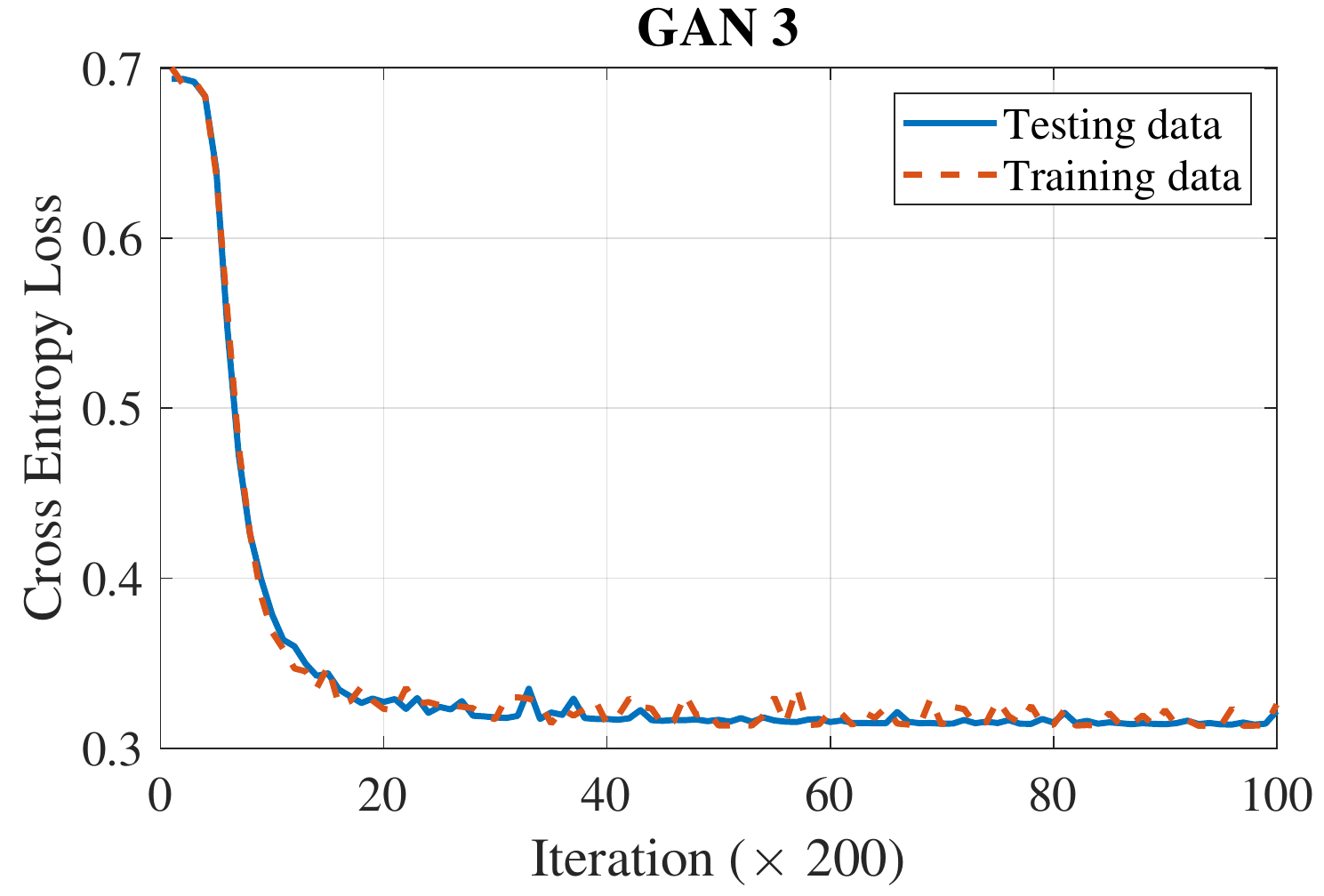}\par 
    \end{multicols}
\begin{multicols}{3}
    \includegraphics[width=\linewidth]{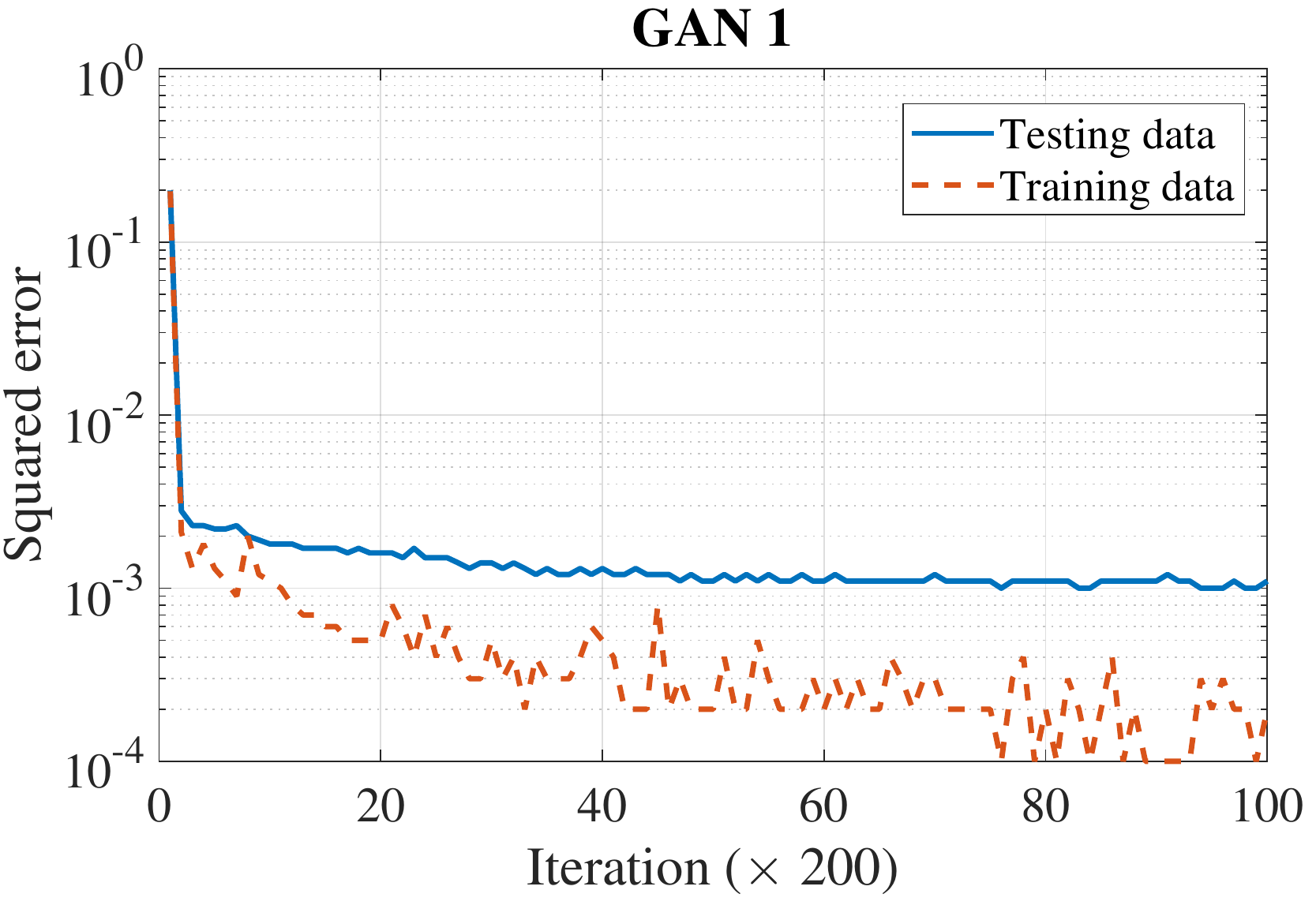}\par
    \includegraphics[width=\linewidth]{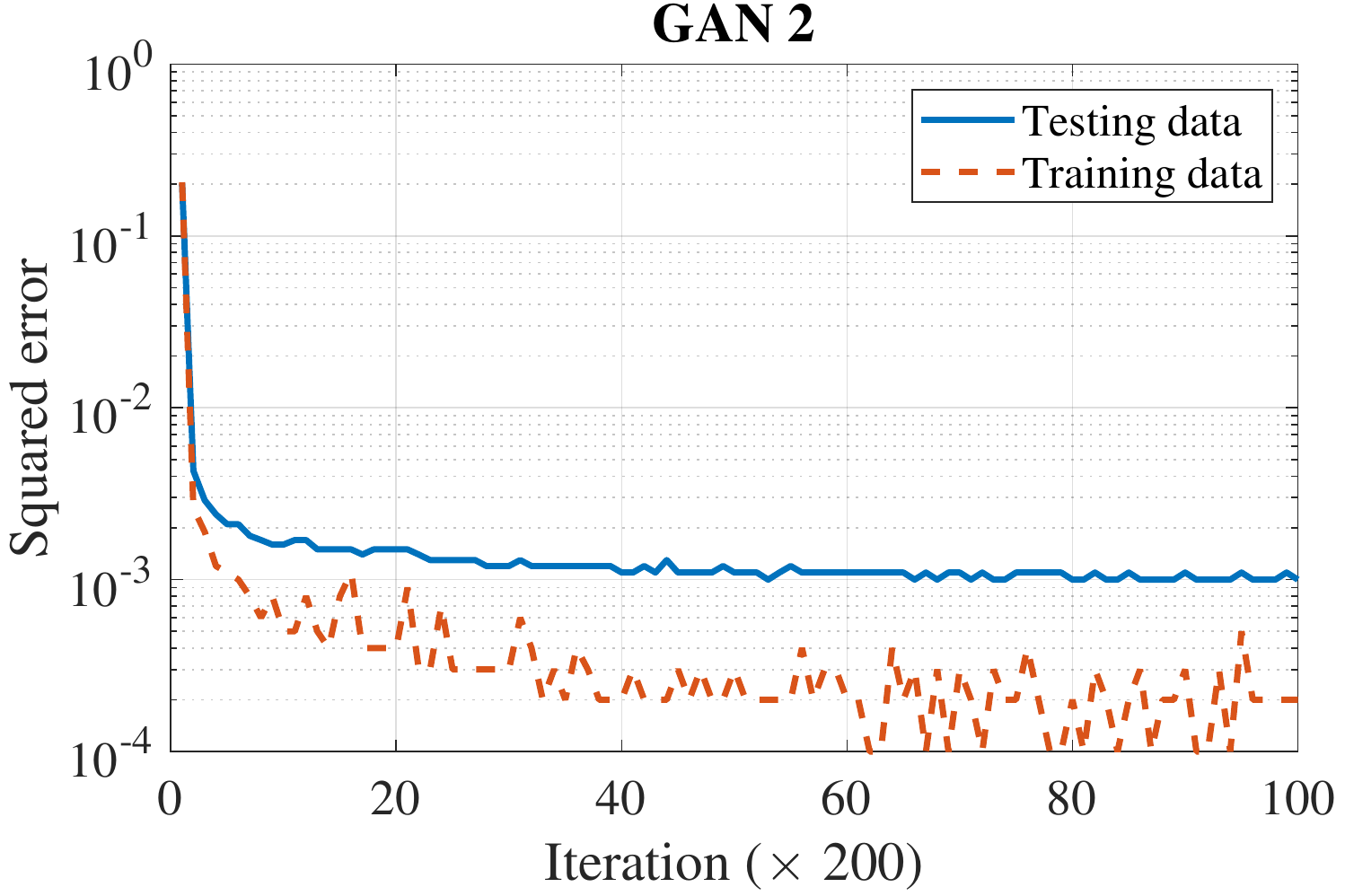}\par
    \includegraphics[width=\linewidth]{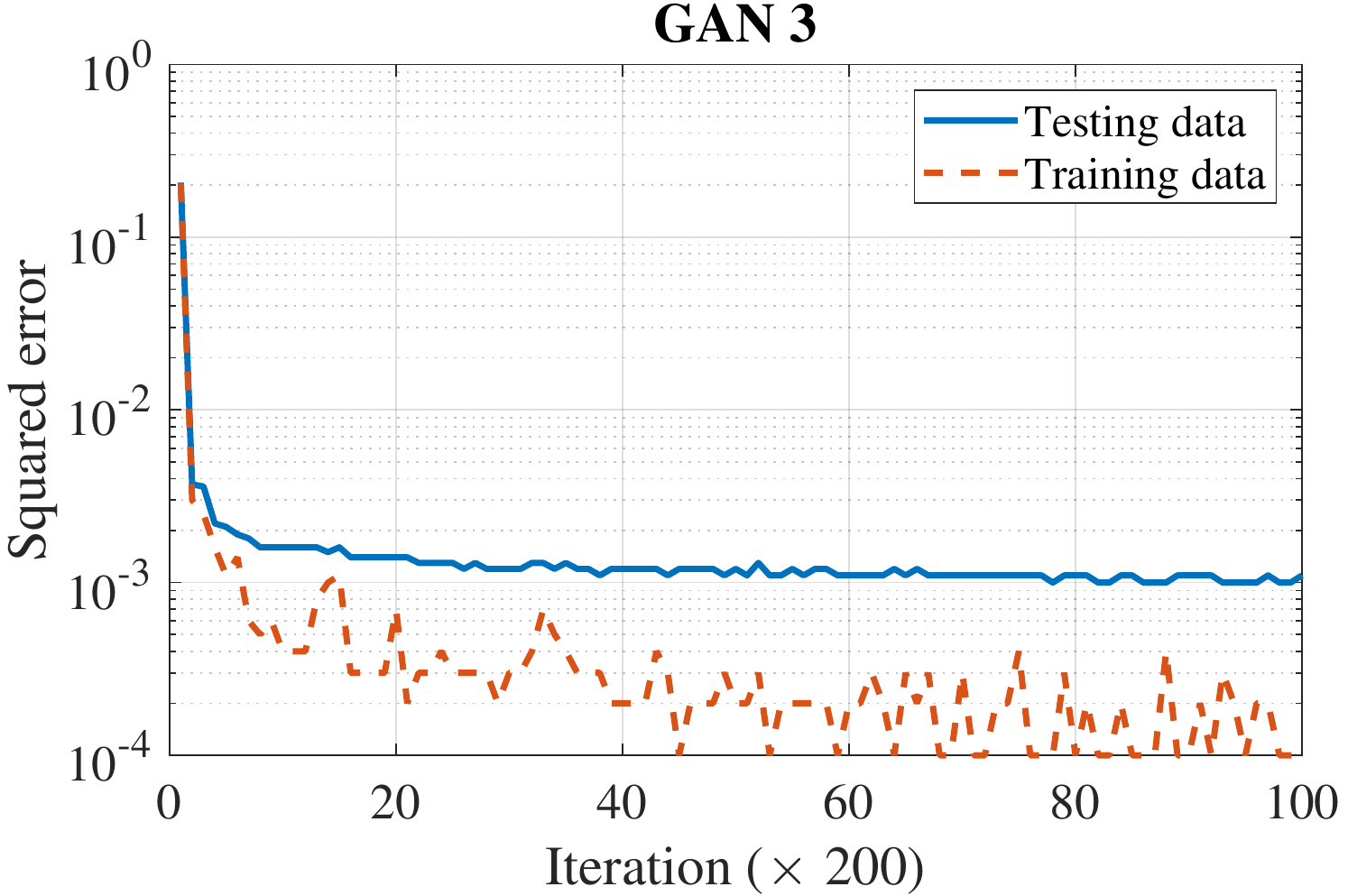}\par 
\end{multicols} \vspace{-4mm}
\caption{The cross-entropy loss and the squared error for each of the three GANs in the HGAN-based model.}\label{HGAN result} \vspace{-4mm}
\end{figure*}

To demonstrate the learning ability of the HGAN-based model, the cross-entropy loss and the squared error of each sub-GAN in~(\ref{g loss in HGAN}) are depicted in Fig.~\ref{HGAN result}. 
To check for overfitting, the testing data is used to test the model in each training step. 
Note that the testing data are not used to train the HGAN-based model. As seen from Fig.~\ref{HGAN result}, for all three GANs, the cross-entropy loss and the squared error dramatically decrease at the beginning of the training phase and converge after that. For instance, the cross-entropy loss for training data in \textit{GAN} 1 drops from 0.7 to 0.33, and the squared error of the training data decreases from 0.1965 to 0.0002. It is also found that with the predicted data from lower GANs, the cross-entropy loss of \textit{GAN} 2 and 3 drops faster than the loss in \textit{GAN} 1. The faster drop indicates that combining the predicted and the measured data leads to more distinguishable transient features for the GAN model, and thus the learning speed is increased.
The  cross-entropy loss and the squared error reduction indicate that the developed model is effectively trained to improve the classification and prediction accuracy. Comparisons of the training and testing loss and error demonstrate that no overfitting occurs. This is concluded by the fact that the cross-entropy loss of the testing data does not rise as the loss of the training data decreases. In other words, both these values maintain a consistent difference. 

\begin{figure}
    \centering
    \includegraphics[width =0.95 \columnwidth]{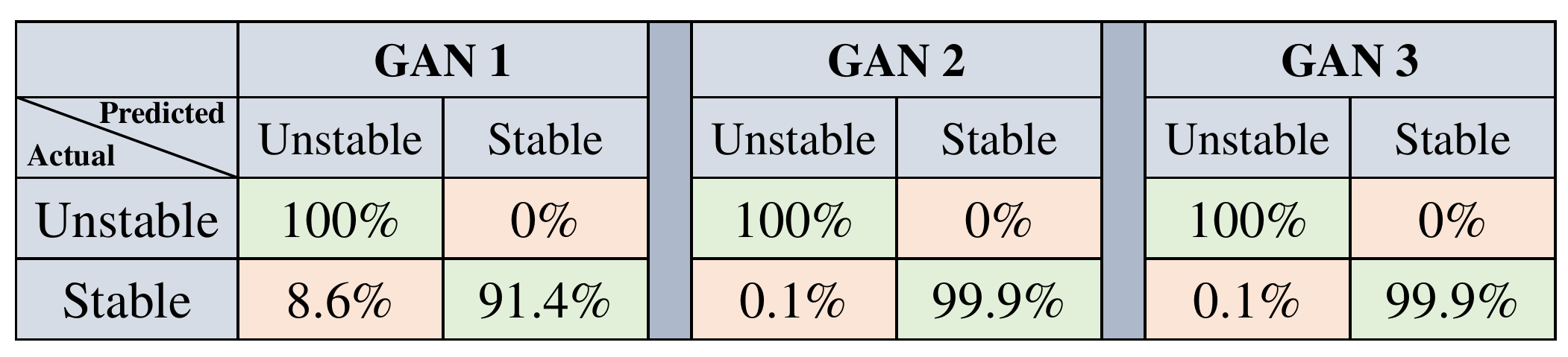}
    \caption{The confusion matrix of each GAN in the HGAN-based TSA model} \vspace{-6mm}
    \label{confusion matrix}
\end{figure}

Upon training the developed HGAN model, the testing data, i.e., 1000 stable and 1000 unstable events, are used to evaluate the classification accuracy of each GAN. The classification results are presented in Table~\ref{HGAN accuracy}. With the single measured PMU voltage data, the classification accuracy of the lowest level \textit{GAN 1} is 95.35\%. However, combining the real measured data and the predicted data from the lower level GANs yields a classification accuracy of 99.95\% for both \textit{GAN} 2 and \textit{GAN} 3. 
Since a single sampled data carries limited transient information, the classification accuracy of \textit{GAN 1} is only 95.3\%. With more predicted data that hold the spatial and temporal features of a transient event, the hidden oscillation patterns of the event are more apparent to the classifiers and result in higher classification accuracy. 
The confusion matrix of each GAN is depicted in Fig.~\ref{confusion matrix}. For \textit{GAN 1}, it can be observed that all unstable transient data are correctly identified, while 8.6\% of the stable events are incorrectly classified as unstable. Benefiting from the predicted PMU voltage data from \textit{GAN 1}, in \textit{GAN 2}, the percentage of misclassified stable events drops to 0.1\%. 
Combining the classification results from these three GANs yields an accuracy of 99.95\% for TSA.
The computation time for TSA using the developed model with three stacked GANs is 0.00298s, which is a reasonable time for near real-time analysis. 

\begin{table}[]
\centering
\caption{TSA classification accuracy of each GAN in the HGAN-based TSA}
\label{HGAN accuracy}
\resizebox{0.7\columnwidth}{!}{%
\begin{tabular}{c|c|c|c}
\hline \hline
 & GAN 1 & GAN 2 & GAN 3 \\ \hline
Accuracy & 95.35 \% & 99.95 \% & 99.95 \% \\ \hline
\end{tabular}%
} \vspace{-2mm}
\end{table}

\begin{figure}[]
    \centering
    \includegraphics[width=\columnwidth]{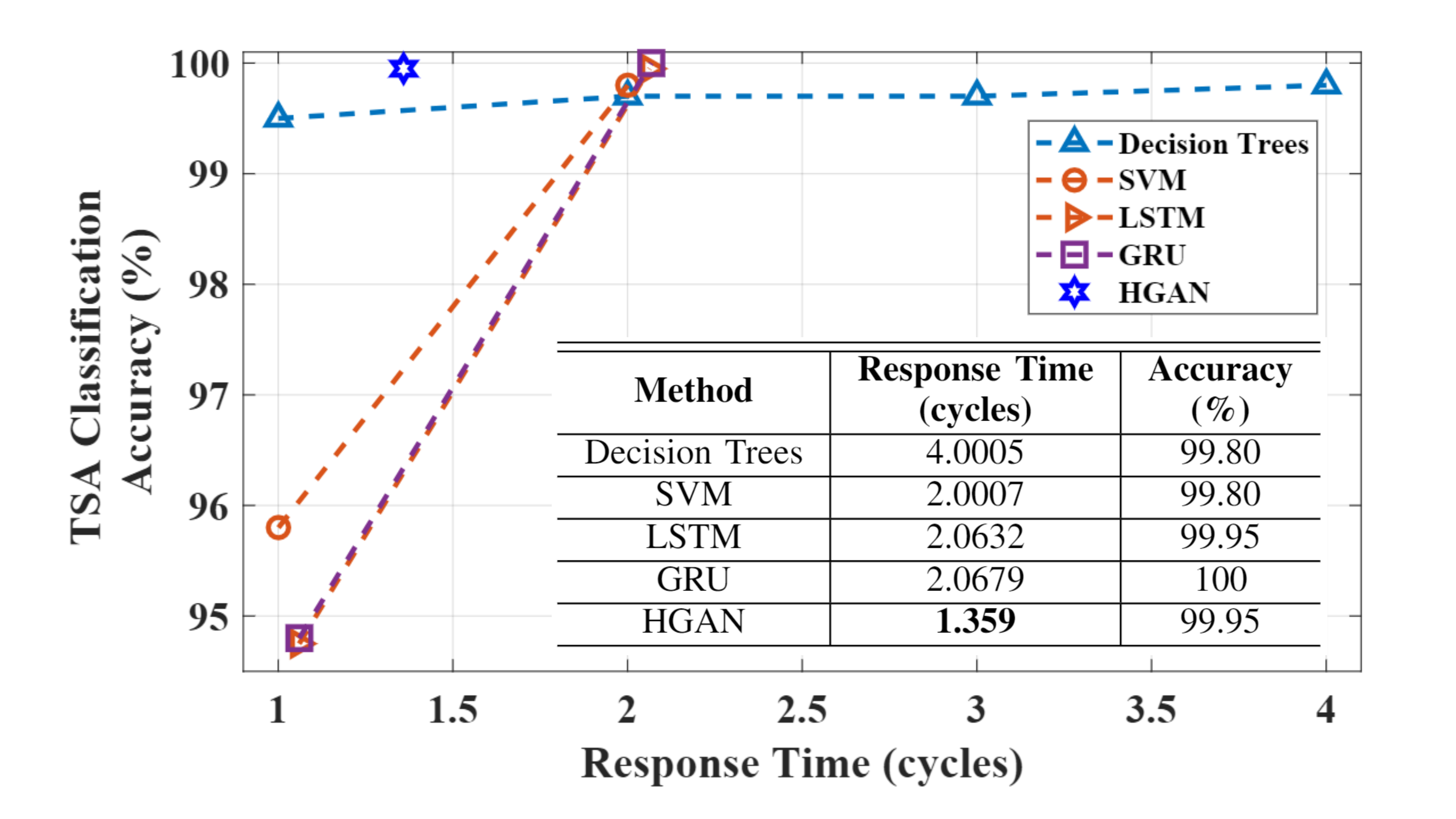}
    \caption{Comparison of response time between the baseline methods and HGAN-based TSA method. In total, 20 PMUs are used in the IEEE 118-bus system. To achieve the same accuracy as the HGAN-based method, other methods would use more samples of PMU measurements and thus require longer response time.}\vspace{-2mm}
    \label{response_time}
\end{figure}

\vspace{1mm}
\noindent \textbf{---Comparison with Baseline Methods}:
To further demonstrate the effectiveness of the developed data-driven TSA approach, its performance is compared with the conventional data-driven baseline methods that deploy other machine learning techniques, namely, decision trees, SVM, LSTM, and GRU. The inputs to the baseline methods are the measured post-fault PMU voltage data. To achieve the highest classification accuracy, the LSTM hyperparameters, namely, the learning rate, training iteration, number of layers, hidden layers, and batch size are set to 0.0005, 40000, 2, and 40, and 128 respectively. Similarly, with various trials on GRU, the optimal learning rate, training iteration, number of layers, and hidden layers are found to be 0.0005, 40000, 2, and 50, respectively.
The comparisons between the baseline methods and the HGAN-based TSA are presented in Table~\ref{baseline} and Fig.~\ref{response_time}.
Compared to SVM, LSTM, and GRU, the decision trees and the HGAN-based TSA achieve higher classification accuracy, i.e., 99.5\% and 99.95\%, respectively with only one sample of PMU measurements. On the other hand, to reach a similar TSA classification accuracy, the HGAN-based method requires less response time, i.e., 1.359 cycles. 
Although the SVM, GRU, and LSTM based-TSA methods can achieve the similar classification accuracy with around 2 cycles, the reduction of 0.65 cycles achieved by the developed HGAN-based method is critical for online TSA. The transient instability issues propagate rapidly in the grids. Hence, a faster TSA allows more time for mitigative actions to prevent further severe consequences.
The response time is the combination of the waiting time for sufficient PMU samples and the computation time of the TSA method.
Although the decision tree-based TSA achieves a relatively high classification accuracy with one sample data, decision trees do not have the ability to predict the transients, while the developed HGAN-based model predicts the transients. This predictive capability makes the developed method suitable for other applications, e.g., cascading failure analysis and power system planning, that would need transient analysis. Hence, the HGAN-based model outperforms the other four baseline methods from both the accuracy and applicability perspectives.

\begin{table}[]
\centering
\caption{TSA classification accuracy of the baseline methods and the HGAN-based TSA with only one sample of PMU measurements (20 PMUs are used)}
\label{baseline}
\resizebox{0.7\columnwidth}{!}{%
\begin{tabular}{c|c}
\hline \hline
\textbf{Method} & \textbf{TSA Classification Accuracy} \\ \hline
Decision tree & \textbf{99.5\%} \\ \hline
SVM & 95.8\% \\ \hline
LSTM & 94.75\% \\ \hline
GRU & 94.8\% \\ \hline
HGAN & \textbf{99.95\%} \\ \hline
\end{tabular}%
}\vspace{-1mm}
\end{table}


\subsection{Impact of PMU Measurement Variations}
\begin{figure}
    \centering
    \includegraphics[width=0.95\columnwidth]{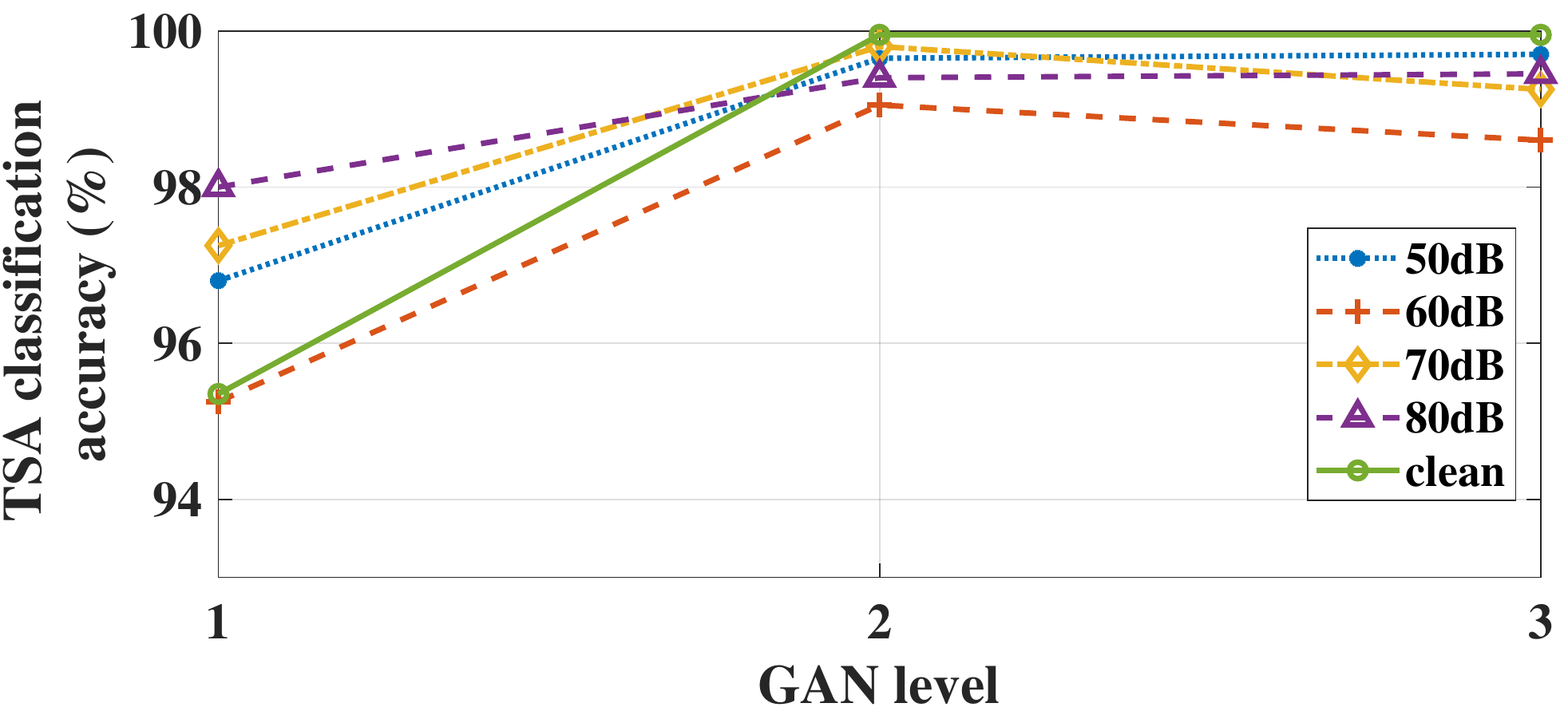}
    \caption{The TSA classification accuracy of each sub-GAN under different noise conditions. The testing data are used for evaluation.} \vspace{-3mm}
    \label{noise impact}
\end{figure}

The PMU measurements used in this study are obtained from simulations and are thus noise free. However, in practice, PMU measurements are contaminated by noise. Hence, the sensitivities of the HGAN-based TSA method to noisy PMU measurements are analyzed. Following the study in~\cite{al2018measurement},
and~\cite{liu2018data}, white Gaussian noise is added to the simulated PMU data. Four different signal-to-noise ratios (SNR), namely, 50dB, 60dB, 70dB, and 80dB, are considered. The classification accuracies of the HGAN-based method under different noise conditions are presented in Fig.~\ref{noise impact}. The classification results in Fig.~\ref{noise impact} show that the noise mainly affects the performance of \textit{GAN 1}. Larger noise will lead to a lower classification accuracy in \textit{GAN 1}. However, it is observed that the classification accuracy of \textit{GAN 2} and \textit{GAN 3} are above 99\%, and are not impacted by the addition of noise. 
Another interesting observation is that among the five different noise conditions, \textit{GAN 1} achieves the highest classification accuracy when a noise of 80dB SNR is added to the measurements. This accuracy is even higher than using clean PMU data. This observation can be explained by the known fact that adding a small noise to the input data leads to a better performance of the GAN models, as they learn more general features of the transient data for a better classification performance~\cite{noh2017regularizing}.

In the existing power systems, PMUs are not deployed on every bus. Hence, depending on where the PMUs are located, the performance of the HGAN-based TSA could vary. To evaluate the performance of the developed approach under varying PMU placement scenarios, five scenarios of different PMU placements are investigated. The PMU locations for these five scenarios are listed in Table~\ref{location PMU impact}. The classification accuracy for each GAN under these five scenarios is depicted in Table~\ref{location impact}. Results in Table~\ref{location impact} show that the location of PMUs significantly affects the classification performance for \textit{GAN 1}, where the accuracy ranges from 94.55\% to 97.9\%. Benefiting from the predicted data, the impact of the location is decreased in \textit{GAN 2} and \textit{GAN 3}, and the classification accuracy is improved. It can also be observed that the classification accuracy slightly drops in \textit{GAN 3}, due to the accumulation of small prediction errors in lower GANs.

\begin{table}[]
\centering
\caption{PMU location scenarios for investigating the impact of PMU placement on the developed TSA approach}
\label{location PMU impact}
\resizebox{\columnwidth}{!}{%
\begin{tabular}{c|c}
\hline \hline
\textbf{Scenario} & \textbf{Location} \\ \hline
1 & 5, 12, 15, 17, 37, 49, 59, 69, 71, 77, 80, 85, 92, 96, 100, 105 \\ 
2 & 5, 12, 15, 17, 37, 56, 67, 69, 70, 71, 77, 80, 92, 96, 100, 105 \\ 
3 & 5, 12, 15, 17, 37, 49, 67, 69, 70, 71, 77, 80, 92, 96, 100, 105 \\ 
4 & 5, 12, 15, 17, 32, 37, 49, 56, 59, 71, 77, 80, 85, 92, 96, 100 \\ 
5 & 5, 12, 15, 17, 32, 37, 49, 56, 59, 69, 71, 77, 80, 92, 96, 100 \\ \hline
\end{tabular}%
}\vspace{-1mm}
\end{table}

\begin{table}[]
\centering
\caption{The TSA classification accuracy of each GAN with different PMU locations. The testing data are used for evaluation. A total number of 16 PMUs are used}
\label{location impact}
\resizebox{0.95\columnwidth}{!}{%
\begin{tabular}{c|c|c|c}
\hline \hline
\multirow{2}{*}{\textbf{PMU Location Scenario}} & \multicolumn{3}{c}{\textbf{TSA Classification Accuracy (\%)}} \\ \cline{2-4} 
 & \multicolumn{1}{c|}{\textbf{GAN 1}} & \multicolumn{1}{c|}{\textbf{GAN 2}} & \multicolumn{1}{c}{\textbf{GAN 3}} \\ \hline
\textbf{1} & 94.55 & 98.55 & 97.90 \\ \hline
\textbf{2} & 97.15 & 99.40 & 98.40 \\ \hline
\textbf{3} & 96.65 & 99.65 & 99.15 \\ \hline
\textbf{4} & 97.90 & 98.55 & 97.55 \\ \hline
\textbf{5} & 95.40 & 98.25 & 97.20 \\ \hline
\end{tabular}%
}\vspace{-2mm}
\end{table}
\begin{table}[b]
\centering
\caption{TSA classification accuracy of each sub-GAN with different number of PMUs}
\label{number PMU}
\resizebox{0.8\columnwidth}{!}{%
\begin{tabular}{c|c|c|c}
\hline \hline
\textbf{Number of PMUs}& \textbf{GAN 1} & \textbf{GAN 2} & \textbf{GAN 3}\\ \hline
\textbf{6} & 86.90\% & 84.65\% & 88.20\%  \\ \hline
\textbf{8} & 91.20\% & 91.15\% & 77.95\%  \\ \hline
\textbf{10} & 94.20\% & 94.55\% & 95.20\%  \\ \hline
\textbf{12} & 95.75\% & 99.60\% & 99.65\% \\ \hline
\textbf{14} & 96.05\% & 99.93\% & 99.92\%   \\ \hline
\textbf{16} & 96.65\% & 99.65\% & 99.15\%  \\ \hline
\textbf{18} & 96.25\% & 99.40\% & 99.91\%  \\ \hline
\textbf{20} & 95.35\% & 99.95\% & 99.95\% \\ \hline
\end{tabular}%
}\vspace{-3mm}
\end{table}
The impact of the number of PMUs on the HGAN-based TSA is also investigated. The classification results for different numbers of PMUs are given in Table~\ref{number PMU}. It can be observed that increasing the number of PMUs improves the TSA accuracy for \textit{GAN 1}. 
When the number of PMUs is larger than 12, the classification accuracy of \textit{GAN 2} and 3 are around 99\%. However, with less than 10 PMUs the TSA classification accuracy of \textit{GAN 2} and \textit{GAN 3} drops, indicating that the predicted data from the lower GANs are not close to the real ones. The prediction error propagates to higher GANs and leads to a large cumulative error, which drastically degrades the classification accuracy. For example, with 8 PMUs, the classification accuracy is 91.15\% in \textit{GAN 2} and drops to 77.95\% in \textit{GAN 3} due to the propagation of the error.
Hence, it can be concluded that more PMU data improve the classification accuracy. However, when the number of PMUs is larger than a threshold, e.g., 12 in the IEEE 118-bus system, the classification accuracy remains fixed in higher level GANs.

\subsection{Discussions}
\begin{figure}[]
        \centering
        \begin{subfigure}[b]{0.47\columnwidth}
            \centering
            \includegraphics[width=1\columnwidth]{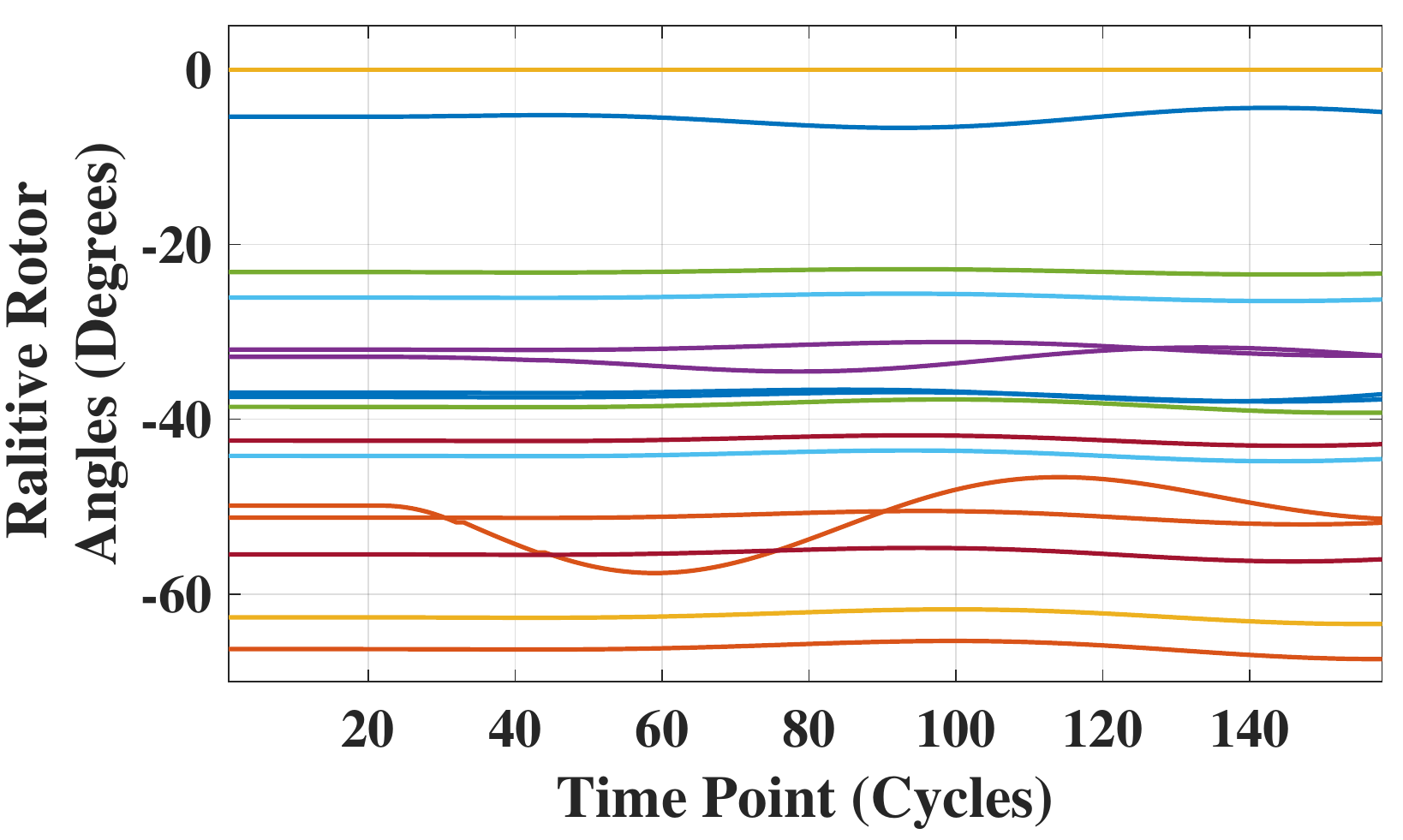}
            \caption[]%
            {Relative rotor angles of a real stable transient}    
            \label{angle stable}
        \end{subfigure}
        \hfill
        \begin{subfigure}[b]{0.48\columnwidth}  
            \centering 
            \includegraphics[width=1\columnwidth]{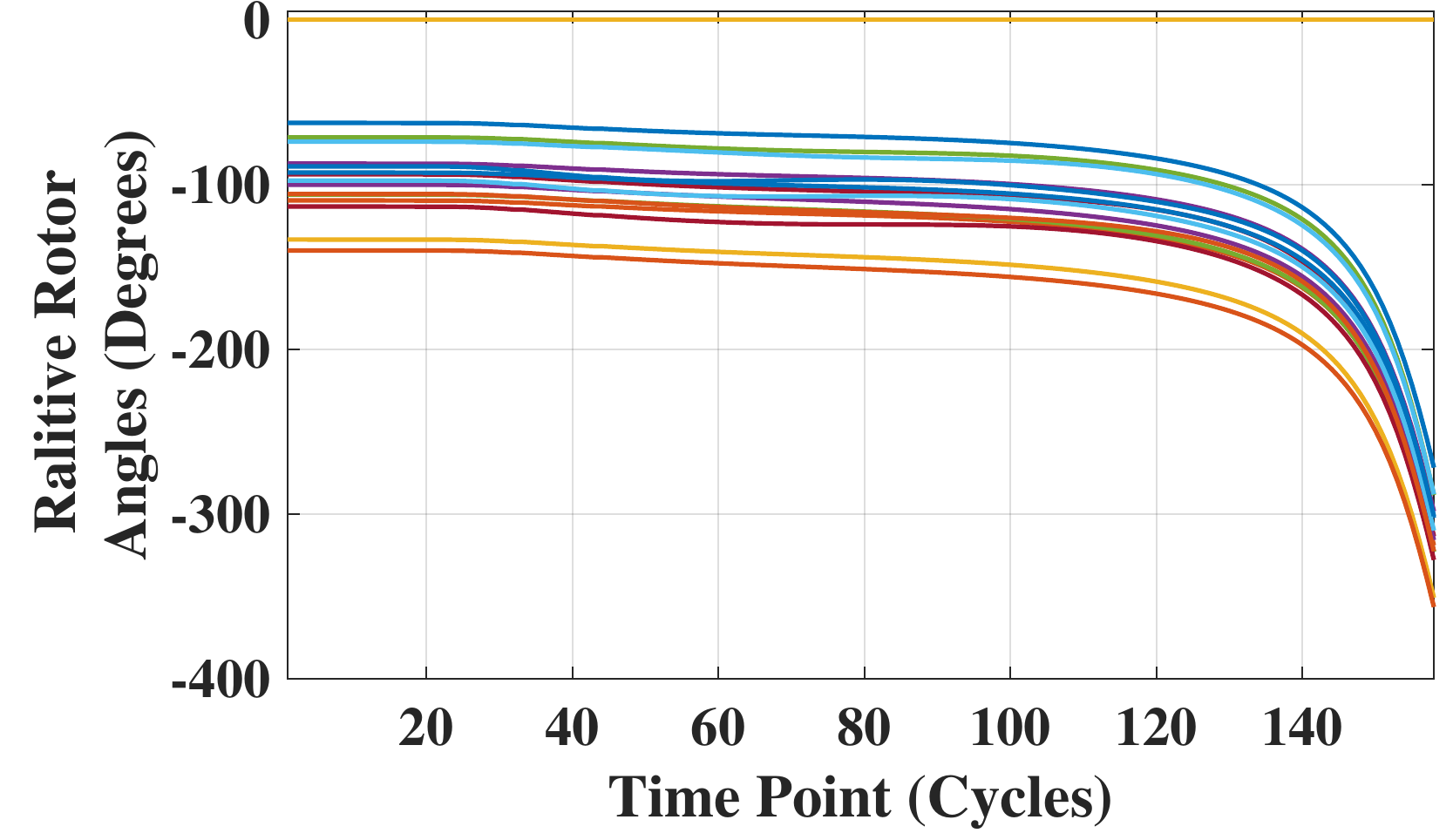}
            \caption[]%
            {Relative rotor angles of a real unstable transient}    
            \label{angle unstable}
        \end{subfigure}
        \vskip\baselineskip
        \begin{subfigure}[b]{0.475\columnwidth}
            \centering
            \includegraphics[width=1\columnwidth]{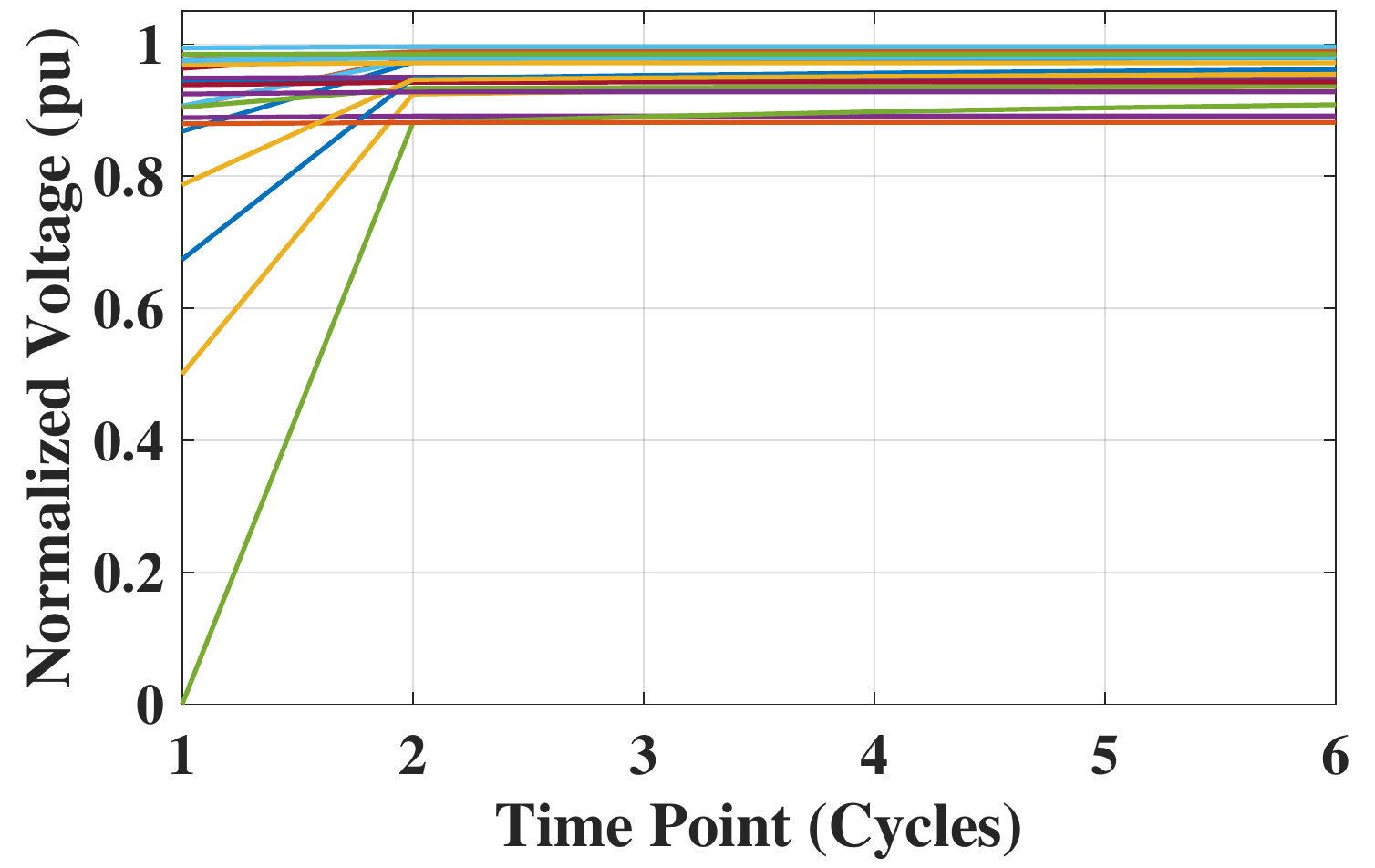}
            \caption[]%
            {Real stable transient}    
            \label{real s}
        \end{subfigure}
        \hfill
        \begin{subfigure}[b]{0.475\columnwidth}  
            \centering 
            \includegraphics[width=1\columnwidth]{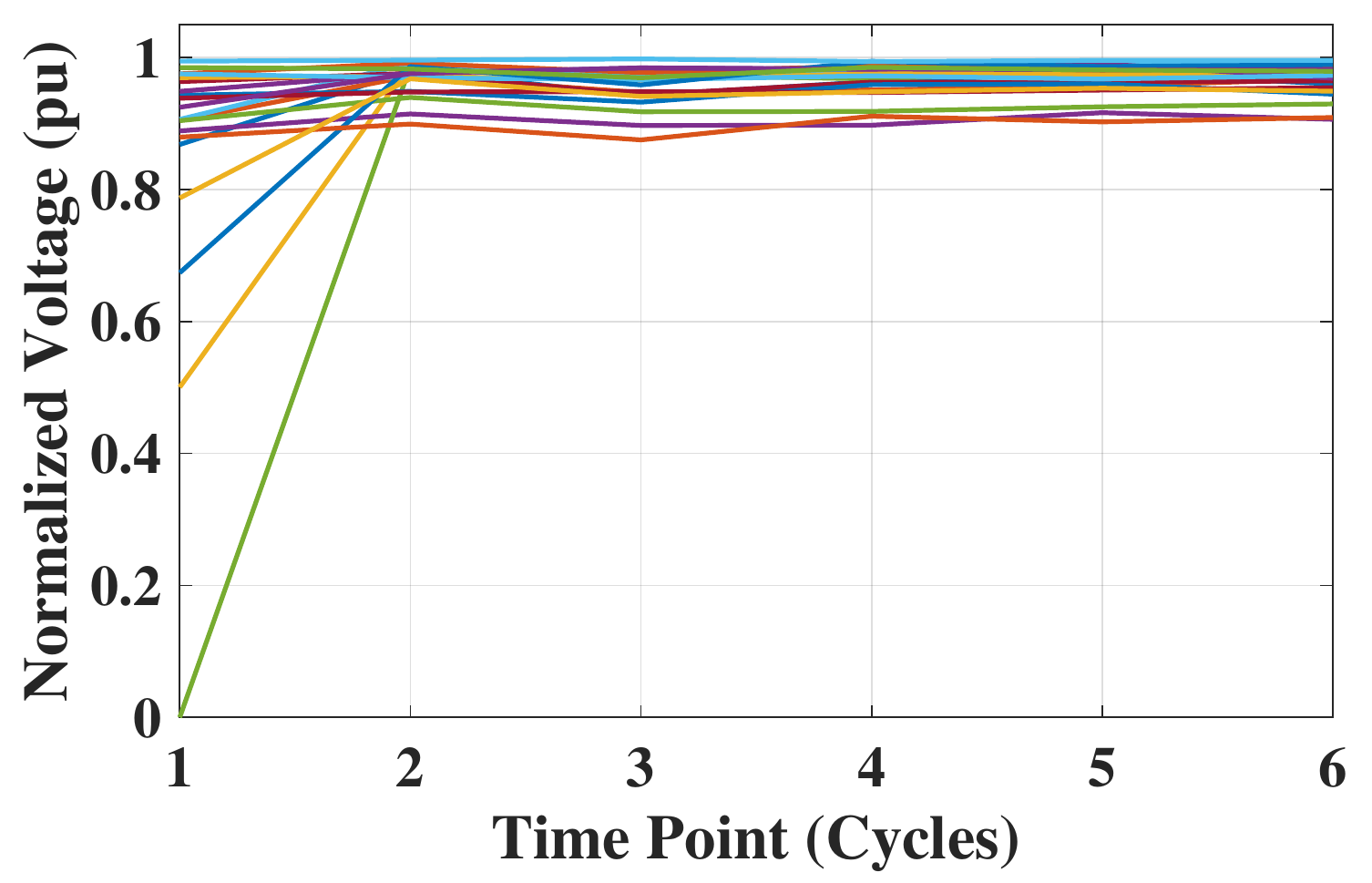}
            \caption[]%
            {Predicted stable transient}    
            \label{fake s}
        \end{subfigure}
        \vskip\baselineskip
        \begin{subfigure}[b]{0.475\columnwidth}   
            \centering 
            \includegraphics[width=\columnwidth]{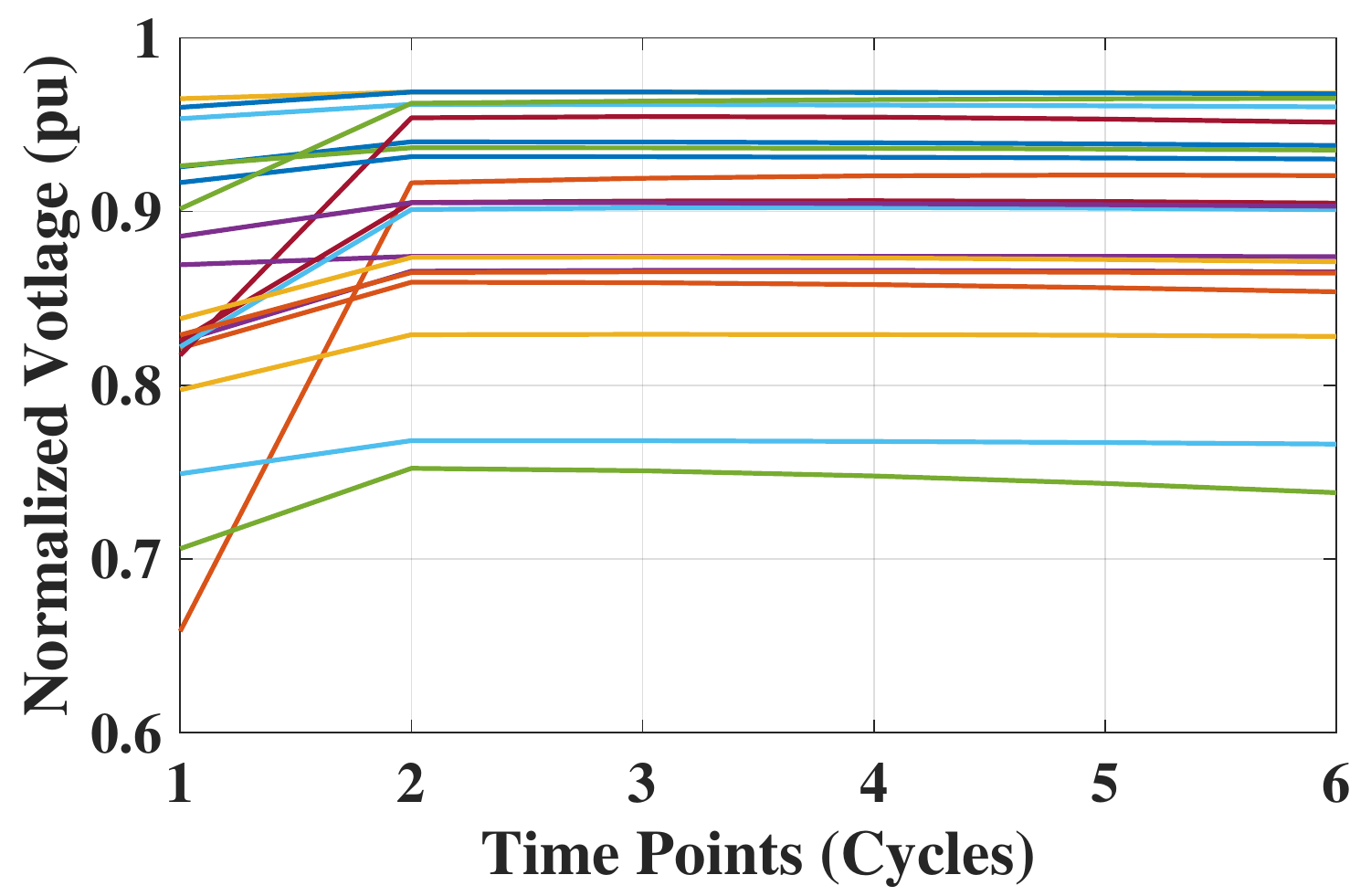}
            \caption[]%
            {Real unstable transient}    
            \label{real un}
        \end{subfigure}
        \hfill
        \begin{subfigure}[b]{0.475\columnwidth}   
            \centering 
            \includegraphics[width=\columnwidth]{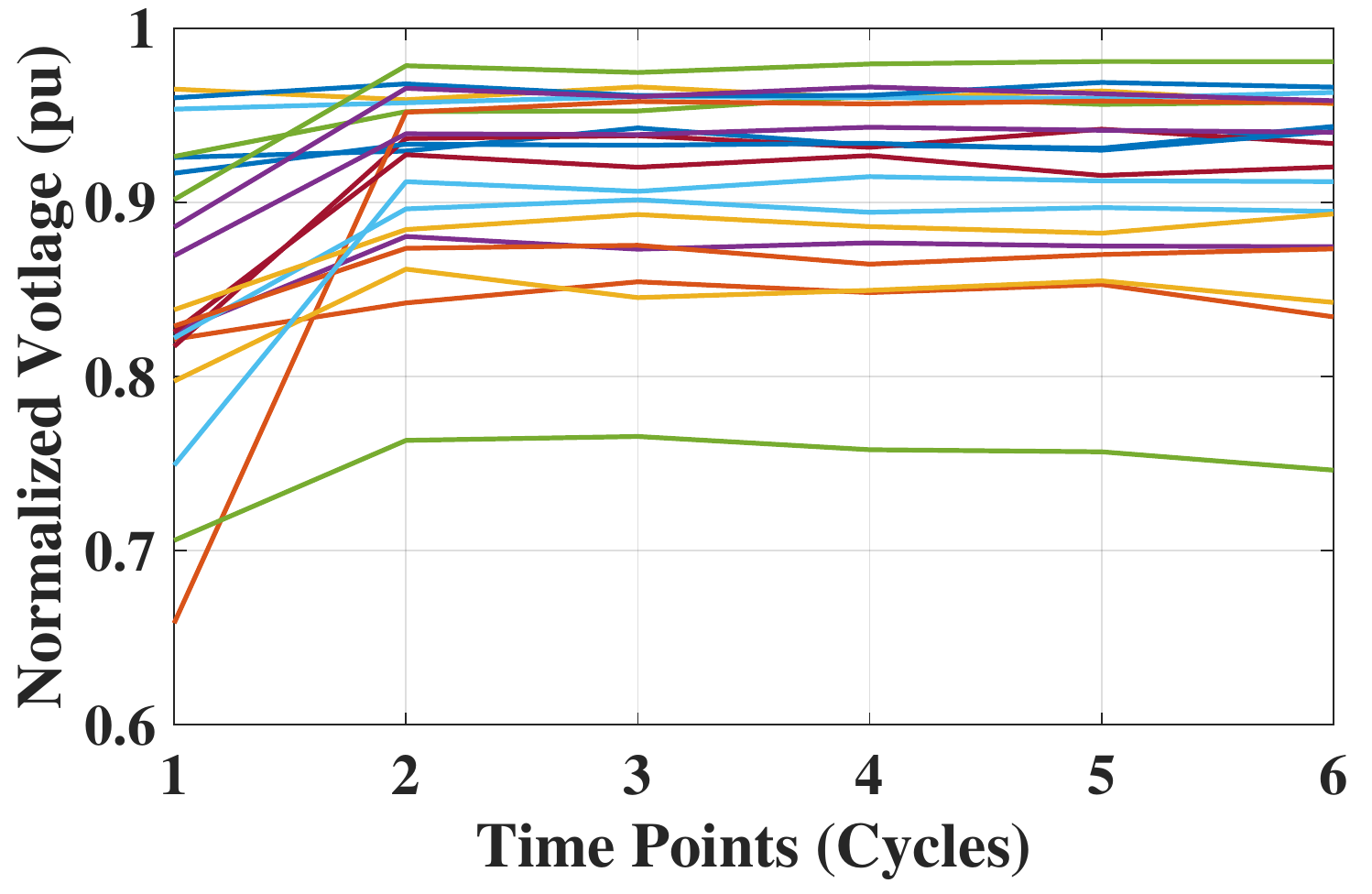}
            \caption[]%
            {Predicted unstable transient}    
            \label{fake un}
        \end{subfigure}
        \caption[ ]
        {Comparison between the real and predicted post-fault stable and unstable transients. The developed TSA approach has 5 GANs and the number of PMUs is 20. The PMU voltage measurements are normalized using a Min-Max scaling method.}\vspace{-3mm} 
        \label{prediction discussion}
\end{figure}

Here, the potential applications of the developed HGAN-based TSA approach, its limitations, and possible solutions are discussed. 
The developed HGAN-based TSA utilizes the predicted transient data to determine the transient stability status of a system. With the specific design of the generator and the GRU network, the predicted voltage time series data retain the spatial and temporal features of the real data. Examples of the predicted post-fault voltage time series for both stable and unstable transients are given in Fig.~\ref{prediction discussion}. 
The relative rotor angles of the stable and unstable transient events are shown in Fig.~\ref{angle stable} and Fig.~\ref{angle unstable}, respectively. 
The HGAN-based TSA has five GANs, and thus can predict the voltage of the subsequent five time steps $\{\hat{v}(2),\cdots,\hat{v}(6)\}$, using the measured post-fault voltage data $v(1)$. It can be observed in Figs.~\ref{real s} and~\ref{real un} that the time series patterns of the two types of real events are different. Compared with the real transient events, only using the measured PMU voltage data from the first time instant, i.e., $v(1)$, the HGAN-based model predicts the oscillation features of these two events, and for the most part, generates similar time series data as the real measurements. 
With the prediction capability, the developed HGAN-based approach is not solely limited to TSA, but can also be deployed in other applications (e.g., cascading failure analysis) that can benefit from the ability to predict how transient events progress.

Although the developed TSA approach is promising for transient analysis, its performance relies on the training data, like any other data-driven model. Here, as the transient data are generated from simulations, the training dataset is ensured to be balanced, i.e, the number of stable and unstable transients is close. 
However, the real-world transient training datasets are imbalanced, i.e., the ratio between the unstable and stable events is less than 5\%. This imbalance occurs since the unstable transient events are high-impact but low-frequency events. An imbalanced training dataset may adversely affect the classification accuracy. To address the problem of imbalanced data, two strategies, i.e., data resampling and redesigning the classifier algorithm, are suggested to improve the HGAN-based TSA method. The data resampling methods, such as oversampling and undersampling methods, can rebalance the ratio between the two types of events~\cite{haixiang2017learning}. In addition to resampling the training data, the classifier algorithm can be adjusted, such that the learning process is more sensitive to the correct identification of the unstable events. For instance, in the cost-sensitive method introduced in~\cite{krawczyk2014cost}, the misclassification cost is modified such that a higher cost is assigned to the wrong prediction of the unstable events.



\section{Conclusion}
This study developed a novel GAN-based TSA approach to promptly and accurately assess the post-fault transient status of a power system using PMU-provided voltage measurements. Unlike the conventional GAN model, the generator in this study is redesigned, such that the modified GAN model predicts the sequence data and learns the distribution of the real data. Benefiting from the specific hierarchical structure of multiple GANs, the developed HGAN-based model maintains the spatial and temporal features of the multivariate PMU time series data. Therefore, with only one sample of PMU measurement, the developed TSA approach improves the  stability assessment accuracy. 
Case studies conducted on the IEEE 118-bus system demonstrate that, with the help of the predicted data, the TSA accuracy can reach 99.95\%. Compared with the other four machine-learning-based methods, i.e., decision trees, SVM, GRU, and LSTM, the HGAN-based TSA achieves a higher classification accuracy and a shorter response time. 
Additionally, to demonstrate the robustness of the HGAN-based TSA, the impacts of measurement noise, location, and the number of available measurements are investigated. It is observed that the prediction capability of the HGAN-based TSA approach makes it robust to these variations and helps improve the classification accuracy. Following proper offline training, the developed TSA can be applied in near real-time to assess the post-fault stability of a power system. Additionally, as the HGAN model enables learning the spatial and temporal features of the system transients, the predicted sequence data can be used for various applications that benefit from the ability to predict the transient response of a system.

In the future, the classifier embedded in the generator can be redesigned to be robust to imbalanced training datasets. The encoder and decoder can also be added to the generator to further improve the predicted unstable transient data.


\bibliographystyle{IEEEtran}
\bibliography{mypaper.bib}

\end{document}